\pdfoutput=1
\documentclass[aps,pre,twocolumn,superscriptaddress]{revtex4-2}

\usepackage{amsmath}
\usepackage{amssymb}
\usepackage{amsfonts}
\usepackage{graphicx}
\usepackage{color}
\usepackage{hyperref}

\begin{document}


\title{Chaotic Diffusion of Dissipative Solitons:\\From Anti-Persistent Random Walk to Hidden Markov Models}


\author{Tony Albers}
\email{tony.albers@physik.tu-chemnitz.de}
\affiliation{Institute of Physics, Chemnitz University of Technology, 09107 Chemnitz, Germany}
\author{Jaime Cisternas}
\affiliation{Complex Systems Group, Facultad de Ingenier\'{\i}a y Ciencias Aplicadas, Universidad de los Andes, Monse\~nor Alvaro del Portillo 12455, Las Condes, Santiago, Chile}
\author{G\"unter Radons}
\affiliation{Institute of Physics, Chemnitz University of Technology, 09107 Chemnitz, Germany}
\affiliation{Institute of Mechatronics, 09126 Chemnitz, Germany}


\date{\today}

\begin{abstract}
In previous publications, we showed that the incremental process of the chaotic diffusion of dissipative solitons in a prototypical complex Ginzburg-Landau equation, known, e.g., from nonlinear optics,
is governed by a simple Markov process leading to an Anti-Persistent Random Walk of motion or by a more complex Hidden Markov Model with continuous output densities.
In this article, we reveal the transition between these two models by studying the soliton dynamics in dependence on the main bifurcation parameter of the Ginzburg-Landau equation
and identify the underlying hidden Markov processes.
These models capture the non-trivial decay of correlations in jump widths and symbol sequences representing the soliton motion, the statistics of anti-persistent walk episodes, and the multimodal density of the jump widths.
We demonstrate that there exists a physically meaningful reduction of the dynamics of an infinite-dimensional deterministic system to one of a probabilistic finite state machine
and provide a deeper understanding of the soliton dynamics under parameter variation of the underlying nonlinear dynamics.
\end{abstract}


\maketitle

\section{\label{sec:I}Introduction}

Deterministic chaotic diffusion, normal or anomalous, is a phenomenon well-known in nonlinear dynamics for at least four decades,
first from low-dimensional Hamiltonian dynamics arising in plasma and accelerator physics \cite{chirikov1979,lichtenberg1992}, later also from solid state physics problems \cite{zacherl1986,geisel1987}.
In the presence of dissipation, the relevant dimension of such problems may be further reduced, which led to studies of diffusion arising in one-dimensional iterated maps \cite{geisel1982,fujisaka1982,geisel1984,geisel1985}.
There, the mechanisms for chaotic diffusion are easily understood.
Deterministic chaotic motion in such low-dimensional systems is often described by a stochastic model.
A prominent example from standard textbooks \cite{lichtenberg1992,ott1993} is the kicked rotor (also known as Standard Map) \cite{chirikov1979},
where for a large enough kick strength, the chaotic motion can be described by a diffusion equation, whose diffusion coefficient depends on the kick strength in a complicated way that can only be determined from simulation data.
Such a connection between deterministic systems and stochastic models would also be desirable for higher-dimensional systems because recently,
chaotic diffusion was also found to occur in non-random infinite-dimensional dynamical systems such as delay systems \cite{wischert1994,schanz2003,sprott2007,lei2011,dao2013_1,dao2013_2,albers2022},
lattice dynamical systems \cite{nishiura2005,egorov2013}, and systems whose dynamics is ruled by partial differential equations (PDEs) \cite{cisternas2016,cisternas2018}.
In contrast to the above mentioned low-dimensional systems, the mechanisms leading to deterministic chaotic diffusion in these PDEs are not understood at all.
The system in question is the cubic-quintic complex Ginzburg-Landau equation, the generic amplitude equation for spatially extended systems near subcritical bifurcations \cite{saarloos1992,aranson2002}.
It is of relevance in diverse fields such as nonlinear optics \cite{AkhmedievAnkiewicz}, binary fluid convection \cite{KBS88}, surface reactions \cite{RJVE91},
and the Faraday instability in granular media \cite{UMS96} and colloidal suspensions \cite{LHARF99}.
The diffusing objects in these systems are dissipative solitons, which may change position due to repeated ‘explosions’ \cite{SAA00,AST01,CCDB12,gurevich2019}.
Such explosions were already found experimentally in passively mode-locked lasers \cite{GA12,RBE15,RBE16,LLY16,YLK18}.
Furthermore, they were also found in time-delayed systems \cite{schelte2019} because of their general relation to partial differential equations \cite{yanchuk2017}.
So, the diffusing objects in these systems are already of more complicated nature, solitary waves with chaotically varying shape, whose center of mass is the quantity of interest.
In one spatial dimension, this is just a scalar quantity, which can be regarded as a projection from infinite-dimensional state space onto the spatial coordinate.
Recently, we showed for one specific case in one dimension that the center of mass can be described phenomenologically as an Anti-Persistent Random Walk \cite{albers2019_1}.
In general, however, even in one spatial dimension, the situation is much more complex.
In a further recent publication \cite{albers2019_2}, we showed for another specific case that the incremental process for the soliton motion is more generally well described by a Hidden Markov Model (HMM) \cite{dymarski2011}.
These models are best known from speech recognition tasks \cite{rabiner1989,indurkhya2010}, but they are rather general statistical models \cite{elliott1995}, also known as probabilistic finite state machines \cite{paz1971}.
They can be adapted via machine-learning algorithms to all kinds of data, ranging, e.g., from neural spike trains \cite{radons1994}, ion channel data \cite{becker1994},
and binding-unbinding transitions in the molecular diffusion of proteins \cite{das2009} to financial data \cite{liehr2000}.
Machine learning techniques are also currently successfully applied to nonlinear dynamical systems \cite{lu2017,pathak2017,pathak2018,zimmermann2018}.
In our analysis of the soliton dynamics, we do not rely on machine learning algorithms, because we aim at describing central physical quantities such as correlation functions of the soliton velocity
or the statistics of anti-persistent walk episodes, which are not automatically represented optimally if one optimizes probability distributions over observation sequences as is done, e.g.,
by the Expectation-Maximization (EM) algorithm \cite{dempster1977,michalek1999} most often used for HMM parameter estimation.
As a result of our approach, we are able to directly determine structure and parameters of the hidden Markov process without applying the typical estimation procedures for HMMs.
In this paper, we aim at an understanding of the transition from Anti-Persistent Random Walks to Hidden Markov Models as the main bifurcation parameter of the Ginzburg-Landau equation is varied.
To do so, we uncover the hidden Markov processes underlying the soliton dynamics for several specific choices of the bifurcation parameter.

The paper is organized as follows.
In Sec.~\ref{sec:II}, we recall the definition of the Ginzburg-Landau equation, its properties as well as its numerical solution leading to the study of the soliton dynamics.
In Sec.~\ref{sec:III}, we briefly recapitulate the concepts related to Hidden Markov Models and how they can be applied to diffusing solitons.
In Sec.~\ref{sec:IV}, we give an overview of our findings from previous publications \cite{albers2019_1,albers2019_2} before we present new complementary results in Sec.~\ref{sec:V}.
Our findings are discussed in Sec.~\ref{sec:VI} and summarized in Sec.~\ref{sec:VII}.

\section{\label{sec:II}The cubic-quintic complex Ginzburg-Landau equation}

The system under investigation is the cubic-quintic complex Ginzburg-Landau (CGL) equation
\begin{equation}
\label{eq:CGL}
\partial_tA=D\partial^2_xA+\mu A+\beta|A|^2A+\gamma|A|^4A\,,
\end{equation}
where in nonlinear optics $A(x,t)\in\mathbb{C}$ is the envelope of the electric field and $\mu<0$ accounts for linear losses.
All other parameters are complex numbers, whose real and imaginary parts have a certain meaning within nonlinear optics \cite{AkhmedievAnkiewicz}.
$D=D_r+\text{i}\,D_i$, where $D_r>0$ characterizes spectral filtering and $D_i$ is the group velocity dispersion coefficient.
$\beta=\beta_r+\text{i}\,\beta_i$, where $\beta_r>0$ is the nonlinear gain coefficient and $\beta_i$ accounts for nonlinear dispersion.
$\gamma=\gamma_r+\text{i}\,\gamma_i$, where $\gamma_r<0$ represents the saturation of the nonlinear gain and $\gamma_i$ corresponds to the saturation of the Kerr nonlinearity.
Except for $\mu$, which is our main bifurcation parameter, we fix all other parameters to constant values, $D=0.125+0.5\text{i}$, $\beta=1+0.8\text{i}$, $\gamma=-0.1-0.6\text{i}$.
Because the CGL cannot be solved analytically, we numerically integrated Eq.~(\ref{eq:CGL}) using a split-step Fourier method with a Runge-Kutta method of fourth order for the nonlinear terms.
We considered a spatial domain of size $L=50$ with periodic boundary conditions consisting of $N=1024$ grid points thus leading to a lattice constant of $dx=50/1024\approx0.05$,
which is appropriate to resolve the high frequency oscillations of the exploding dissipative solitons.
We numerically solved Eq.~(\ref{eq:CGL}) on this spatial domain starting with a Gaussian-shaped initial condition with random `hair' in order to obtain different realizations of the soliton dynamics.
We used small time steps $dt=0.001$
\footnote{The used pseudo-spectral method is always stable and there is no risk of high wavenumber instabilities.
It is often used in the literature \cite{AkhmedievAnkiewicz} even with the same discretization.
The chaotic behavior of the solitons is an inherent property of the system and not induced by numerical instabilities.}
and obtained for each considered value of $\mu$ $400$ time series of duration $T=10^5$.
In order to accelerate the simulations, we implemented the numerical algorithm on graphical processing units programmed with the language PyOpenCL \cite{pycuda}.

\begin{figure}
\includegraphics[width=\linewidth]{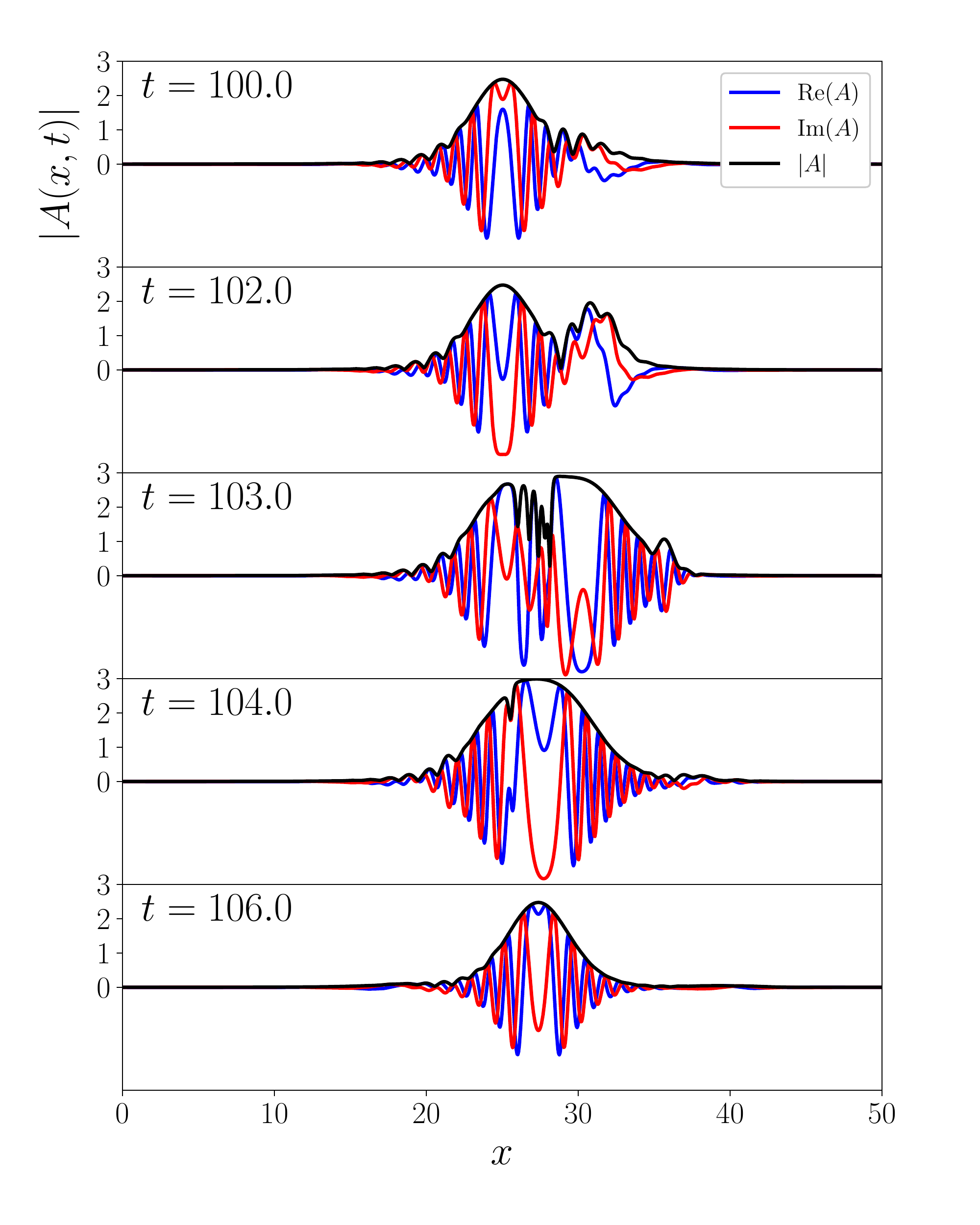}
\caption{\label{fig:explosion}
Explosion of a dissipative soliton for $\mu=-0.10$ showing the time evolution of the real and imaginary part of the complex amplitude $A(x,t)$ as well as its absolute value.}
\end{figure}

We numerically found that dissipative solitons are explosive in the parameter range $\mu_c=-0.21375<\mu<0$.
The time evolution of a single explosion is shown in Fig.~\ref{fig:explosion}.
Before an explosion, the soliton radiates waves to both sides.
Then, seemingly at random, one of these oscillating tails grows, overcomes the original soliton, and then merges with that.
This larger localized solution, however, lasts only for a short time period and then shrinks again to the size of the original soliton.
The solitons before the explosion and after the explosion are almost identical except for a spatial shift because the explosion is asymmetric.
The total duration of such an explosion is roughly 5 units of time, but the interexplosion times as well as the sizes of the spatial shifts can be described statistically due to the chaotic nature of the explosions.
Each numerically generated time series for each value of $\mu$ investigated in this article contains roughly 5200 to 5800 explosions.
A long sequence of such asymmetric explosions leads to a random walk kind of motion of the center of mass of the soliton.
From a dynamical systems point of view, explosions are the result of an enlargement of a chaotic attractor, more specifically, an attractor-merging crisis \cite{cisternas2013}.

\begin{figure}
\includegraphics[width=\linewidth]{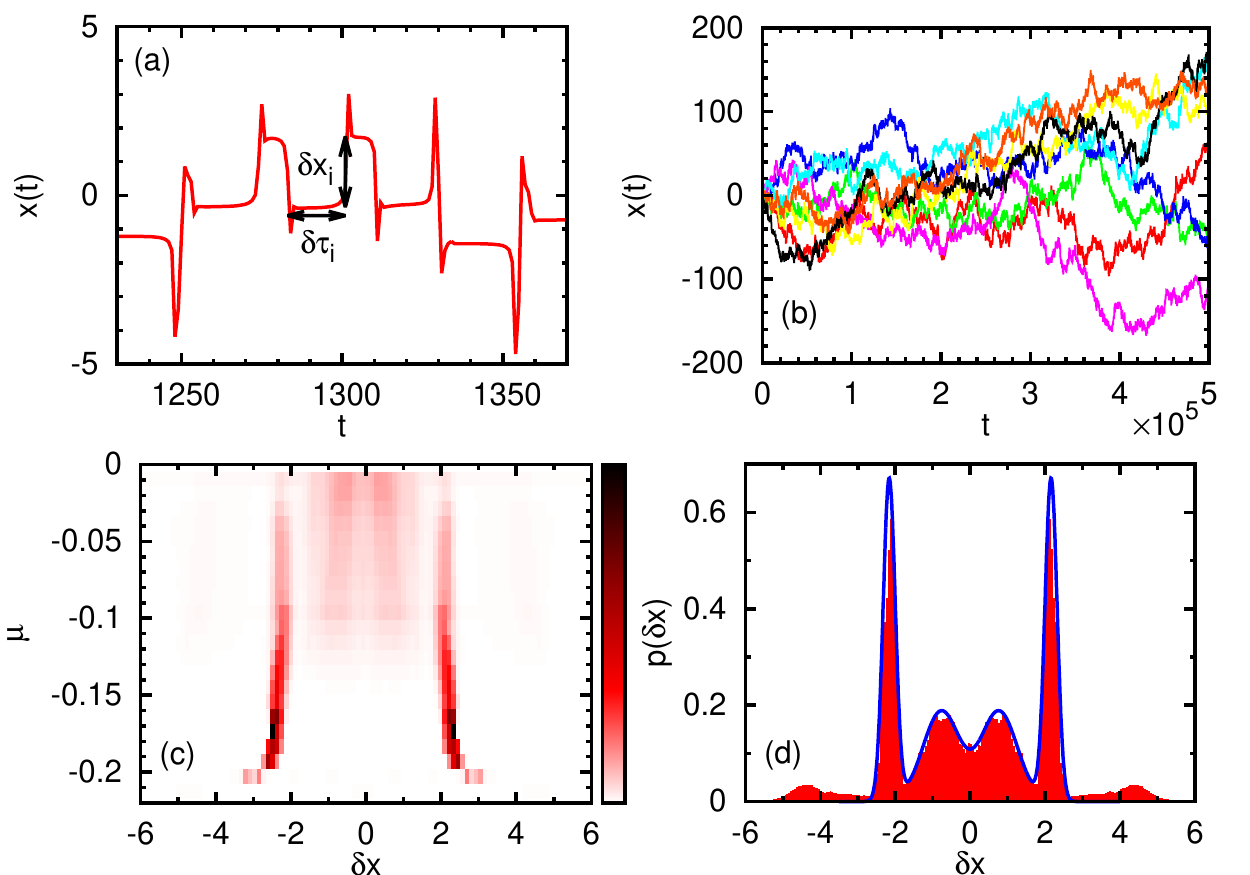}
\caption{\label{fig:trajectories}(color online)
(a) Center of mass $x(t)$ of an explosive dissipative soliton in dependence on time $t$ on a short time scale obtained as numerical solution of Eq.~(\ref{eq:CGL})
for parameters $\mu=-0.10$, $D=0.125+0.5\text{i}$, $\beta=1+0.8\text{i}$, and $\gamma=-0.1-0.6\text{i}$.
Except for $\mu$, these parameters are kept constant through this article.
The basic structure of the soliton trajectory consists of spatial shifts $\delta x_i$ due to asymmetric explosions and inter-explosion times $\delta\tau_i$.
(b) Eight independent realizations of the soliton motion for $\mu=-0.10$ on a large time scale.
Trajectories are reminiscent of Brownian motion.
(c) Distribution $p(\delta x)$ of spatial shifts in dependence on the bifurcation parameter $\mu$ of the CGL depicted as intensity plot.
Dark red means large probability and white means zero probability.
Each horizontal line corresponds to a distribution of spatial shifts for one certain value of $\mu$.
(d) Distribution of spatial shifts for $\mu=-0.10$.
The histogram depicted in red is numerically obtained from the soliton data.
The blue line is a weighted sum of four Gaussian distributions, $0.23\cdot\mathcal{N}(x|\pm0.77,0.24)+0.27\cdot\mathcal{N}(x|\pm2.16,0.026)$,
where $\mathcal{N}(x|u,\sigma^2)$ is a Normal distribution with mean value $u$ and variance $\sigma^2$,
and results from the Hidden Markov Model shown in Fig.~\ref{fig:hmm1} with parameters and emission densities from Eq.~(\ref{eq:parameters1}) and Eq.~(\ref{eq:emissions1}), respectively.
The small humps on the margin of the distribution are neglected for our theoretical consideration.}
\end{figure}

In the following, we reduce the infinite-dimensional dynamics of the CGL to a one-dimensional dynamics by considering the time dependence of the ``center of mass'' of an exploding dissipative soliton.
For a localized solution of Eq.~(\ref{eq:CGL}) on a periodic domain of size $L$, the center of mass can be calculated via
\begin{equation}
\label{eq:center_of_mass}
x(t)=\frac{L}{2\pi}\arg\left(\int_{-L/2}^{L/2}|A(x,t)|^2\exp\left(\frac{2\pi\text{i}x}{L}\right)\,\text{d}x\right)\,,
\end{equation}
which allows us to consider the resulting soliton motion.
Figure~\ref{fig:trajectories} (a) shows the typical structure of a soliton trajectory consisting of almost constant ``plateaus'', which are interrupted by chaotic bursts due to the explosions.
Because the explosions are asymmetric, the plateaus before and after an explosion are in general different.
This kind of motion can be idealized as a sequence of spatial shifts (jumps) $\delta x_i$ and inter-explosion times (waiting times) $\delta\tau_i$, which can be treated statistically due to the chaotic dynamics.
On a large time scale, this sequence leads to an apparently random walk and is reminiscent of Brownian motion, see Fig.~\ref{fig:trajectories} (b).
Moreover, the mean-squared displacement (MSD) of an ensemble of soliton trajectories increases linearly in time (not shown in the figures) indicating normal diffusion.
In order to understand this one-dimensional dynamics in more detail, we investigate the statistics of the inter-explosion times and spatial shifts.
From the random-walk theory it is known that the diffusion coefficient, i.e., the slope of the linear increase of the MSD, is only influenced by the mean of the waiting times and not by the details of their statistics.
Therefore, in the following, we concentrate our analysis on the sequences of spatial shifts.
The distribution of the spatial shifts in dependence on the bifurcation parameter $\mu$ of the CGL is shown in Fig.~\ref{fig:trajectories} (c) as intensity plot.
For small values of $\mu$ in the range $\mu_c<\mu<0$, the distribution consists of only two sharp peaks, but for increasing values of $\mu$, two broader humps with increasing weight appear between them in addition.
Hence, the two sharp peaks lose weight for increasing values of $\mu$.
The distribution of spatial shifts for an intermediate value of the bifurcation parameter, $\mu=-0.10$, is shown in Fig.~\ref{fig:trajectories} (d), where one can see the sharp peaks and the broader humps in between.

\section{\label{sec:III}Hidden Markov Models}

\begin{figure}
\includegraphics[width=\linewidth]{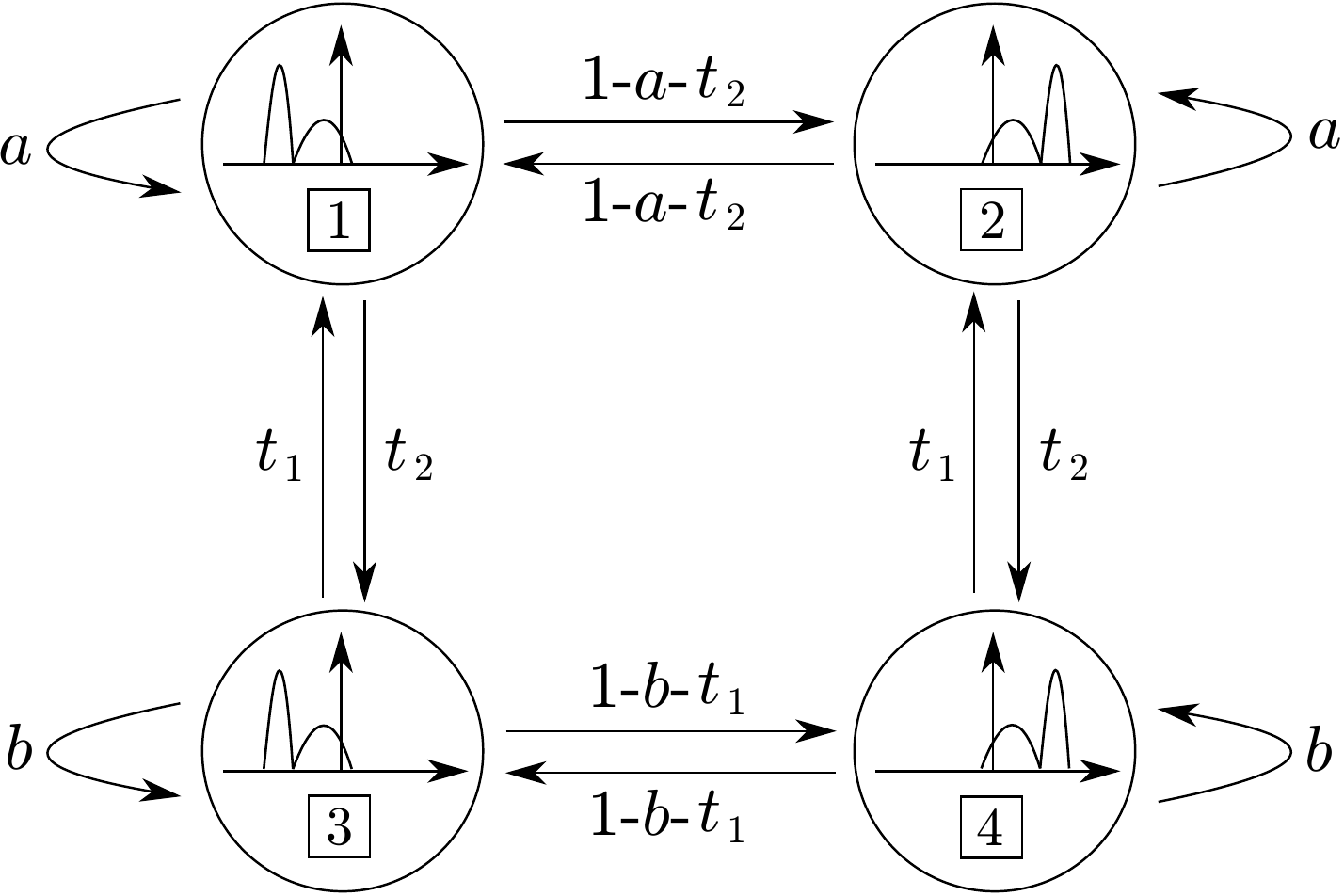}
\caption{\label{fig:hmm}
Schematic illustration of the most general Hidden Markov Model for the dynamics of the spatial shifts $\delta x_t$, which we consider in this article.
It consists of nodes $i=1,2,3,4$ and transitions as indicated.
The curves plotted inside the nodes represent schematically the corresponding output probability densities for the $\delta x_t$.}
\end{figure}

We have seen that the evolution of the system states according to the CGL equation is dominated by the dynamics of solitons, which during repeated explosions, counted by a discrete time index $t$,
undergo spatial shifts $\delta x_t$, $t=1,2,\dots$ which eventually lead to a diffusive behavior of the solitons.
A proper description of the stochastics of the sequences $\{\delta x_t\}$, which has its origin in the chaotic dynamics of the CGL equation, is therefore essential for the understanding of the resulting diffusive motion.
It turns out, as we will show in the sections to follow, that Hidden Markov Models (HMMs) capture very well the stochastic dynamics of the increments $\delta x_t$.
Therefore, in the following, we briefly survey the general principles of Hidden Markov Models.
HMMs are generally used as probabilistic generators of symbol or number sequences of arbitrary length $T$.
Such a sequence is often called an observation sequence $\mathbf{O}=O_1O_2O_3\dots O_T$ with individual observations $O_t$.
In our case, the $O_t$ take real values and correspond either directly to the shifts $\delta x_t$, or, in a simplified setting, to their signs $S_t=\text{sign}(\delta x_t)=\pm1$.
In an abstract sence, an HMM can be regarded as a parametrized probability distribution $P(\mathbf{O}|\Lambda)$,
which is the probability (density) for generating some sequence $\mathbf{O}$ given the model parameters denoted collectively as $\Lambda$.
There are two types of parameters appearing in a typical HMM:
Firstly, there are the parameters defining an underlying discrete state Markov process.
Assuming an $N$-state Markov process, the latter is defined by an $(N\times N)$-transition matrix $\mathbf{T}$ with elements $T_{ij}$
denoting the probability of making in one time step a jump to state $i$ given that one is in state $j$.
Correspondingly, one has $\sum_{i=1}^NT_{ij}=1$, which means that $\mathbf{T}$ is a stochastic matrix containing $N(N-1)$ independent parameters.
In graphical representations of Markov models, arrows are drawn pointing from state $j$ to state $i$ if the corresponding transition probability $T_{ij}$ is non-zero.
It turns out in the following sections that the most general Markov model we have to consider has $N=4$ states, which would imply $12$ independent parameters,
but symmetry considerations and other observations reduce the number of parameters to four, named $a$, $b$, $t_1$ and $t_2$, which determine the transition matrix in the following way
\begin{equation}
\label{eq:transition_matrix}
\mathbf{T}=\begin{pmatrix}a&1-a-t_2&t_1&0\\1-a-t_2&a&0&t_1\\t_2&0&b&1-b-t_1\\0&t_2&1-b-t_1&b\end{pmatrix}\,.
\end{equation}
The corresponding Markov graph consists of the states and arrows shown in Fig.~\ref{fig:hmm}.
Note that in the limit $t_1=t_2=0$, this structure reduces to two independent $2$-state processes, where each of them can be viewed as a generator of an anti-persistent random walk with persistence parameters $a$ and $b$, respectively.

The second ingredient, which turns a Markov model into a Hidden Markov Model, are so-called \textit{emissions}, which generate symbols, or, in our case real numbers, on each state $i$ of the Markov model.
They are drawn randomly from a distribution $\lambda_i(x)$ each time a path in the Markov model visits the state $i$.
That is why HMMs are doubly stochastic processes: paths through the states are generated randomly, and emissions on each state visted are generated randomly.
The distributions $\lambda_i(x)$ can be continuous densities or discrete probability distributions.
In our case, where $x$ represents size and direction of the jumps of the solitons during one explosion, $\lambda_i(x)$ is a continuous density,
whereas its reduction to the direction of the jumps results in discrete distributions with support at $S=\pm1$.
The latter can be written in a continuous density form as $\lambda_i^d(x)=\lambda_i^+\delta(x-1)+\lambda_i^-\delta(x+1)$,
where $\lambda_i^+=1-\lambda_i^-=\int_0^{\infty}\lambda_i(x)\,\text{d}x$ is the probability of a jump in the positive direction given the system is in state $i$.
Correspondingly, the resulting HMM is called Continuous Density HMM (CD-HMM), or, discrete HMM, respectively.
The emissions can be combined with the state transitions by defining a new matrix $\mathbf{T}(x)$ with matrix elements $T_{ij}(x)=\lambda_i(x)T_{ij}$,
or, in the discrete case, with $\lambda_i(x)$ replaced by $\lambda_i^d(x)$.
With these definitions, the probability density $p(x_1,x_2,\dots,x_T|\Lambda)$ for the generation of a sequence of jumps $\delta x_1\delta x_2\dots\delta x_T$ can be written in matrix form as
\begin{equation}
\label{eq:sequence_prob}
p(x_1,x_2,\dots,x_T|\Lambda)=\boldsymbol{\eta}^T\mathbf{T}(x_T)\dots\mathbf{T}(x_2)\mathbf{T}(x_1)\boldsymbol{\pi}_0\,,
\end{equation}
where $\boldsymbol{\eta}^T=(1,1,\dots,1)$ performs the summation over the $N$ components of the vector obtained from iterating the initial distribution $\boldsymbol{\pi}_0$ over the states by the matrices $\mathbf{T}(x_t)$.
In a stationary situation, $\boldsymbol{\pi}_0$ is naturally chosen as the invariant distribution $\mathbf{p}^*$, i.e., $\boldsymbol{\pi}_0=\mathbf{p}^*$ with $\mathbf{p}^*=\mathbf{T}\,\mathbf{p}^*$.
The precise formulation of the meaning of the density in Eq.~(\ref{eq:sequence_prob}) is that $p(x_1,x_2,\dots,x_T|\Lambda)\,\text{d}x_1\,\text{d}x_2\,\dots\,\text{d}x_T$ is the joint probability
of generating by the model $\Lambda$ the first jump value $\delta x_1$ in the interval $[x_1,x_1+\text{d}x_1]$, and the second value $\delta x_2$ in the interval $[x_2,x_2+\text{d}x_2]$, and so on.
To complete the model description, we have to complete $\Lambda$ by the parameters which parametrize the continuous emission densities $\lambda_i(x)$.
It turns out that $\lambda_i(x)$ can be chosen as a mixture of two Gaussian densities, i.e.,
\begin{equation}
\label{eq:Gaussian_mixture}
\lambda_i(x)=\sum_{k=1,2}\phi_{ik}\,\mathcal{N}(x|u_{ik},\sigma_{ik}^2)\,,
\end{equation}
where $\mathcal{N}(x|u,\sigma^2)$ denotes a Gaussian probability density with mean $u$ and variance $\sigma^2$, while the weights $\phi_{ik}\geq0$ obey $\phi_{i1}+\phi_{i2}=1$.
Thus, the parametrization of the $N$ emission densities $\lambda_i(x)$ add in principle $5N$ parameters to the model vector $\Lambda$.
So, for $N=4$, we would have $20$ additional independent parameters, which again through symmetries and observations reduce considerably.
For instance, instead of eight variances, we need only two different variances (broad and narrow).
In Fig.~\ref{fig:hmm}, we schematically show the most general model we have to consider.
Clearly, if one has such a model $\Lambda$, every quantity of interest, e.g., correlation functions, can be calculated,
because the joint probability density in Eq.~(\ref{eq:sequence_prob}) gives the most complete information about the stochastics of the generated sequences.

In typical applications of HMMs, one has data from the outside world or nature, which are observed and represented by an empirical density $p(x_1,x_2,\dots,x_T)$.
Now, one tries to model the data as good as possible, which usually means that the model generated data, characterized by $p(x_1,x_2,\dots,x_T|\Lambda)$, should be as close as possible to the observed data.
Therefore, one usually seeks a parameter set $\Lambda^*$ such that the distance between the distributions $p(x_1,x_2,\dots,x_T)$ and $p(x_1,x_2,\dots,x_T|\Lambda)$, as measured, e.g., by the Kullback-Leibler distance, becomes minimal.
For this task exist powerful iterative algorithms, such as the Expectation-Maximization (EM) algorithm, which generate sequences of models $\Lambda_1\rightarrow\Lambda_2\rightarrow\dots$,
which converge quickly to the, in this sense, optimal model $\Lambda^*$.
Unfortunately, because $p(x_1,x_2,\dots,x_T|\Lambda^*)$ is still an approximation to $p(x_1,x_2,\dots,x_T)$, there is no guarantee that, e.g., correlations in the data, inherent in $p(x_1,x_2,\dots,x_T)$,
are well represented in the data $p(x_1,x_2,\dots,x_T|\Lambda^*)$ generated by such an optimal model.
Since we want to model the diffusional behavior of the observed solitons, it is essential that, in view of the Green-Kubo relation between diffusion and correlation functions, the correlations are captured as good as possible.
Especially, the decay rates in the model $\Lambda$ should properly reproduce the correlation decays in the data.
This, together with the main distributional aspects, is the basis of our determination of optimal parameters $\Lambda^*$.
Such optimal parameters depend naturally on the parameters of the CGL, Eq.~(\ref{eq:CGL}).
In the latter, typically, the gain parameter $\mu$ is varied.
This means that the optimal HMM parameters are functions of $\mu$, i.e., $\Lambda^*=\Lambda^*(\mu)$.
In previous publications, we identified the optimal model parameters $\Lambda^*$ for two cases of the CGL.
In \cite{albers2019_2}, we identified $\Lambda^*(\mu=-0.10)$, which corresponds to a $4$-state HMM as introduced above,
whereas in \cite{albers2019_1}, we found that the diffusive behavior of the solitons for $\mu=-0.18$ is well described by an Anti-Persistent Random Walk (APRW),
whose generator is basically a $2$-state Markov process, i.e., it is not an HMM, or in some sense a degenerate case of an HMM.
Now, one expects that $\Lambda^*(\mu)$ is continuously varying with $\mu$.
This poses the natural question of how a continuous variation of $\mu$ can transform a correspondingly parametrized $4$-state HMM into a $2$-state Markov process.
At first sight, this appears impossible due to the different topological structure of the underlying Markov graphs.
But we will see that a priori, there exist several possibilities, and one goal of this paper is to uncover the mechanism at work in this system.
The answer will be given and discussed in Sec.~\ref{sec:V} and Sec.~\ref{sec:VI}, respectively.
At first, however, we briefly recapitulate some results for the APRW behavior and for the previously found HMM in Sec.~\ref{sec:IV}.

\section{\label{sec:IV}Review of previous results}

\begin{figure}
\includegraphics[width=\linewidth]{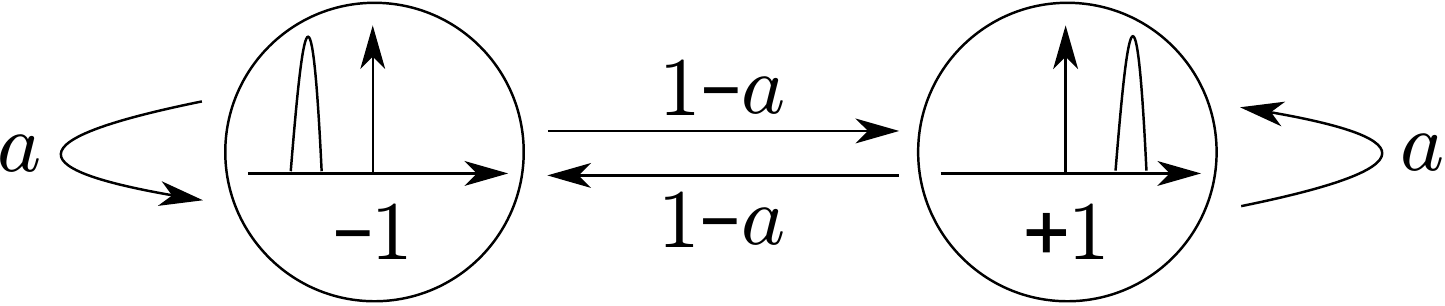}
\caption{\label{fig:aprw}
Schematic representation of a simple $2$-state Markov model for the dynamics of the spatial shifts $\delta x_t$ for $\mu=-0.18$ leading to an Anti-Persistent Random Walk of the soliton motion.}
\end{figure}

In a previous article \cite{albers2019_1}, we investigated the soliton dynamics for $\mu=-0.18$, where according to Fig.~\ref{fig:trajectories} (c) only two sharp peaks are visible in the distribution of the spatial shifts.
We found that the sequence of spatial shifts follows a $2$-state Markov process of first order, see Fig.~\ref{fig:aprw}.
In this model, the sharp peaks belong to two different states representing a spatial shift in the positive or negative direction.
Due to symmetry, these states have the same persistence parameter $a$, which is the probability to remain in the current state during a single time step.
Accordingly, the probability to change the current state, i.e., to change the algebraic sign of the spatial shift, is $1-a$.
In each state, the spatial shift is randomly drawn from the corresponding peaked distribution, which can be approximated by a single Gaussian probability density.
Because we numerically found that the persistence parameter $a$ is very small ($a\approx0.0009$),
the resulting soliton motion is an Anti-Persistent Random Walk (APRW) leading to a mono-exponential decrease of the correlation of spatial shifts.
Note that the $2$-state Markov process in Fig.~\ref{fig:aprw} with emission densities that are single Gaussian distributions can be regarded as a degenerate case of the more general Hidden Markov Model (HMM) in Fig.~\ref{fig:hmm}.

\begin{figure}
\includegraphics[width=\linewidth]{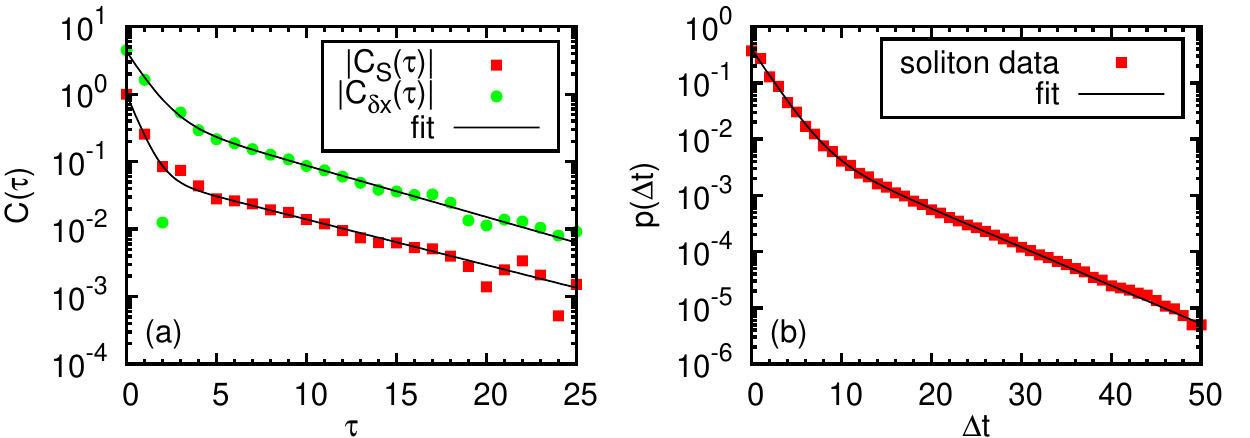}
\caption{\label{fig:statistics1}(color online)
(a) Absolute value of the numerically determined correlations $C_{\delta x}(\tau)$ of spatial shifts $\delta x_t$ (green circles) and $C_S(\tau)$ of the algebraic signs $S_t$ of the spatial shifts (red squares) for $\mu=-0.10$.
The black lines are corresponding bi-exponential fits, $3.80\cdot\exp(-1.03\cdot\tau)+0.51\cdot\exp(-0.18\cdot\tau)$ and $0.95\cdot\exp(-1.61\cdot\tau)+0.066\cdot\exp(-0.16\cdot\tau)$, respectively.
(b) Numerically determined distribution $p(\Delta t)$ of zig-zag streaks for $\mu=-0.10$.
The black line is a bi-exponential fit, $0.39\cdot\exp(-0.56\cdot\Delta t)+0.013\cdot\exp(-0.16\cdot\Delta t)$.}
\end{figure}

In order to understand the implications for the soliton dynamics that are induced by the appearance of the broader humps in the distribution of spatial shifts for larger values of $\mu$,
we investigated the case $\mu=-0.10$ in another recent article \cite{albers2019_2}.
Here, we want to briefly recapitulate our line of arguments leading to a HMM for the dynamics of the spatial shifts:

We start by considering the correlation of the spatial shifts and the correlation of the algebraic signs of the spatial shifts for $\mu=-0.10$.
Figure~\ref{fig:statistics1} (a) shows that both correlations decay bi-exponentially.
This finding, in contrast to the mono-exponential decrease for $\mu=-0.18$, leads to the assumption that the soliton motion might be described by two APRWs with transitions between them.
In order to check this assumption, we consider another statistics, which we call the distribution of zig-zag streaks,
where a zig-zag streak of length $\Delta t$ is given by $\Delta t$ successive changes of the algebraic signs of the spatial shifts, for instance, $+1,-1,+1,-1,-1$ for $\Delta t=3$.
The distribution of zig-zag streaks for $\mu=-0.10$ is shown in Fig.~\ref{fig:statistics1} (b) and decays bi-exponentially too.
This weighted sum of two mono-exponential decays describing the probability of the occurrence of a zig-zag streak of length $\Delta t$ confirms the assumption of a superposition of two APRWs,
because a single APRW shows a mono-exponential decrease of the distribution of zig-zag streaks proportional to $(1-a)^{\Delta t}$, see Fig.~\ref{fig:aprw}.
It is natural to assume that the sharp peaks and the broader humps in the distribution of spatial shifts are connected with two APRWs.
In order to check this, we consider the distribution of zig-zag streaks for two filtered sequences of spatial shifts where the first one contains all $\delta x_t$ with $|\delta x_t|<1.75$
and the second one contains all $\delta x_t$ with $1.75<|\delta x_t|<2.60$.
The resulting sequences belong to the broader humps or the sharp peaks in the distribution of spatial shifts, respectively, see Fig.~\ref{fig:trajectories} (d).
Our numerical results are shown in Fig.~\ref{fig:numerics_theory1} (a).
In the first case, we obtain a mono-exponential decrease with a slope roughly the same as for the first decay in Fig.~\ref{fig:statistics1} (b) and in the second case,
we get a bi-exponential decrease where the slope of the second one is roughly identical to the second exponential decay in Fig.~\ref{fig:statistics1} (b).
Similar results were found for the correlations of the spatial shifts (not shown in the figures).
These observations lead to the conclusion that the sharp peaks contribute to both APRWs, whereas the broader humps only contribute to one of them.

\begin{figure}
\includegraphics[width=\linewidth]{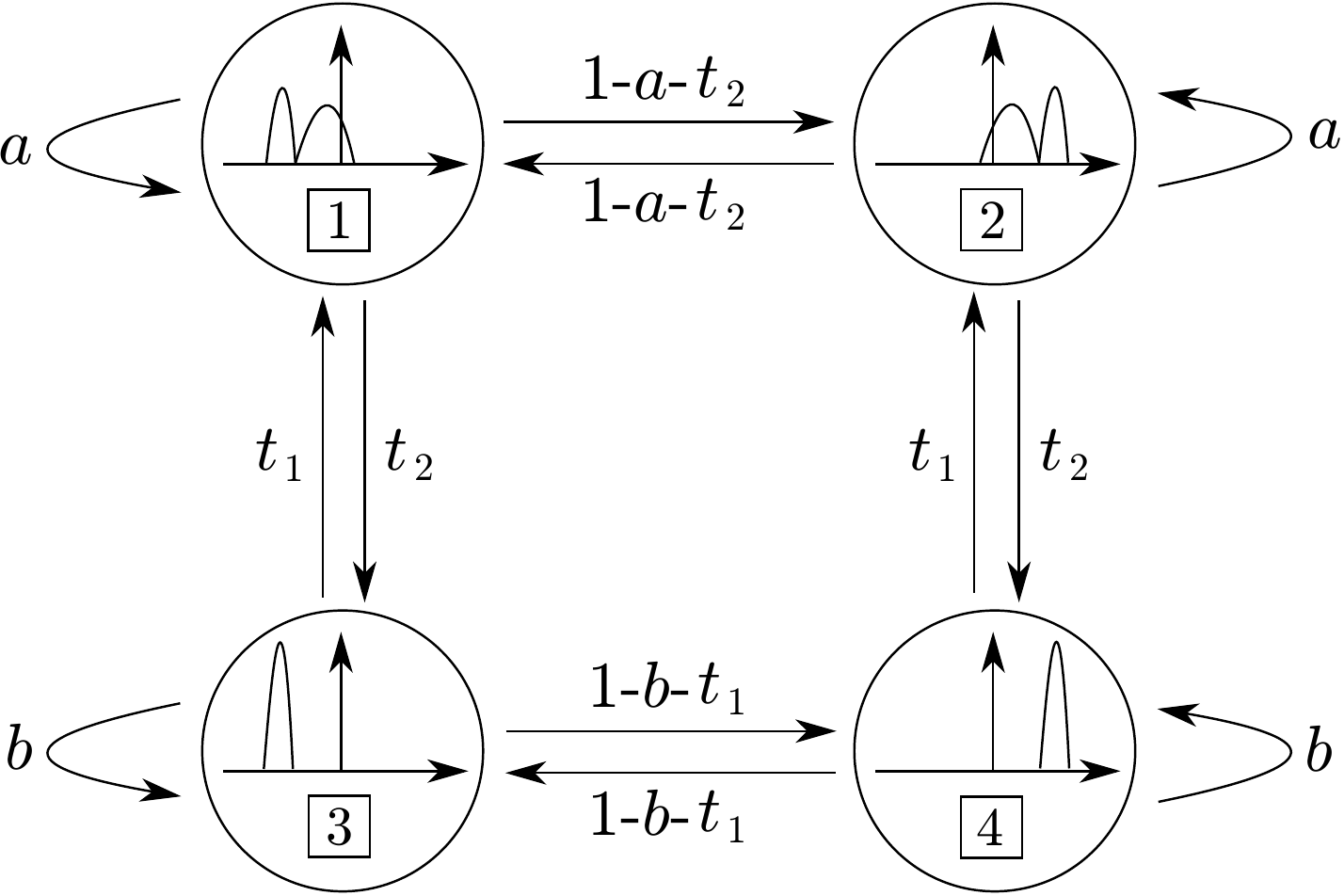}
\caption{\label{fig:hmm1}
Schematic illustration of a Hidden Markov Model for the dynamics of the spatial shifts $\delta x_t$ for $\mu=-0.10$ whose distribution is shown in Fig.~\ref{fig:trajectories} (d).
This model is a special case of Fig.~\ref{fig:hmm} insofar as only one (narrow) Gaussian output density is associated with states $i=3$ and $i=4$, respectively.}
\end{figure}

\begin{figure}
\includegraphics[width=\linewidth]{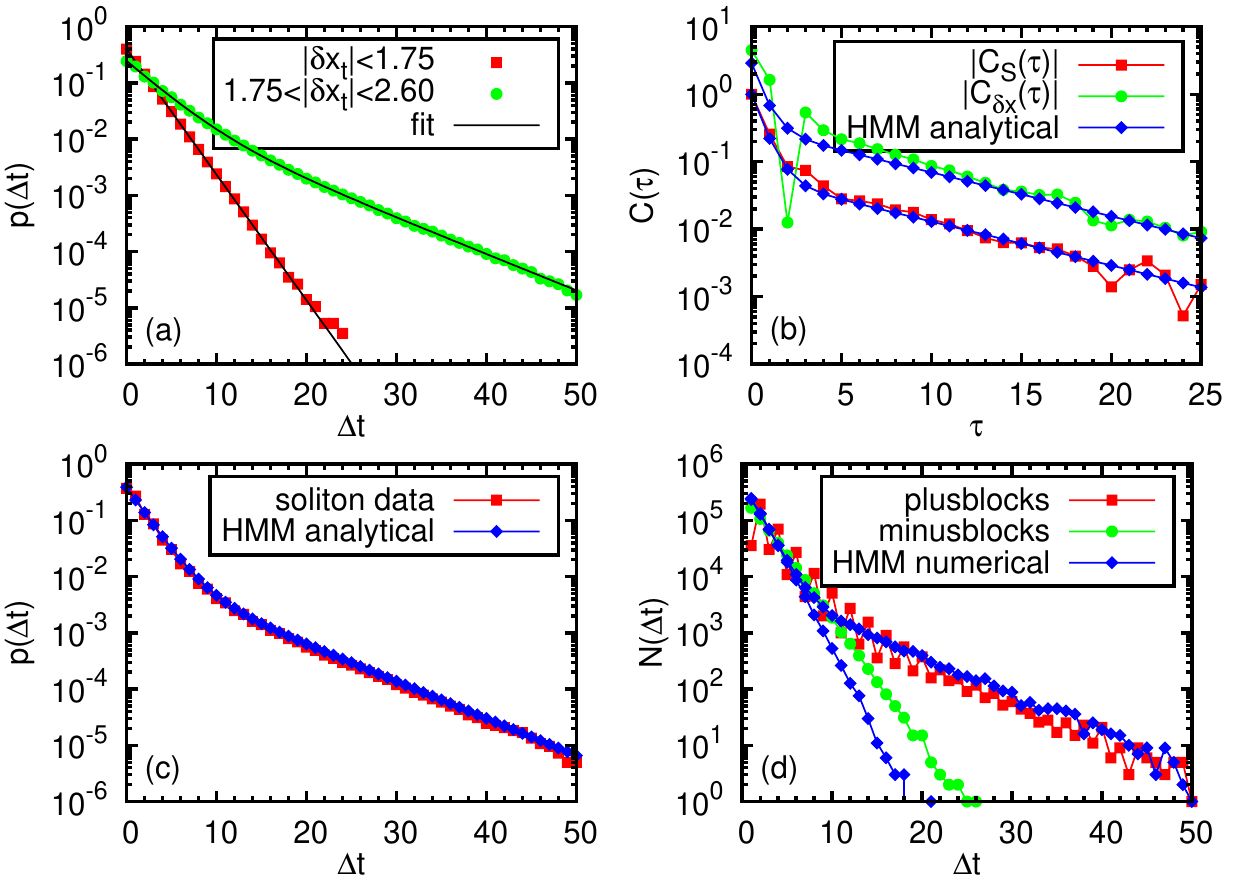}
\caption{\label{fig:numerics_theory1}(color online)
(a) Distribution $p(\Delta t)$ of zig-zag streaks for sequences of spatial shifts where all $\delta x_t$ with $|\delta x_t|>1.75$ (red squares) or all $\delta x_t$ with $|\delta x_t|<1.75$ and $|\delta x_t|>2.60$ (green circles) were deleted.
The black lines are corresponding fits.
In the first case, a mono-exponential decrease is found, $0.40\cdot\exp(-0.51\cdot\Delta t)$, whereas in the second case, a bi-exponential decrease is obtained, $0.21\cdot\exp(-0.34\cdot\Delta t)+0.035\cdot\exp(-0.15\cdot\Delta t)$.
(b) Comparison of the numerically determined correlations $C_{\delta x}(\tau)$ of spatial shifts $\delta x_t$ (green circles) and $C_S(\tau)$ of the algebraic signs $S_t$ of the spatial shifts (red squares) for $\mu=-0.10$
with the corresponding analytical results (blue diamonds) for the HMM shown in Fig.~\ref{fig:hmm1} with parameters and emission densities from Eq.~(\ref{eq:parameters1}) and Eq.~(\ref{eq:emissions1}), respectively.
(c) Numerically determined distribution $p(\Delta t)$ of zig-zag streaks for $\mu=-0.10$ compared with the corresponding analytical result from the HMM.
(d) Absolute frequency for the occurrence of plusblocks and minusblocks (defined in the main text) numerically determined for $\mu=-0.10$ and compared with corresponding numerical results obtained from the HMM.}
\end{figure}

This data analysis leads to the HMM for the dynamics of the spatial shifts shown in Fig.~\ref{fig:hmm1}.
It consists of $N=4$ hidden states $i$, $i=1,2,3,4$, with corresponding emission densities $\lambda_i(x)$.
States $1$ and $2$ represent the dynamics of spatial shifts belonging to the first APRW (with persistence parameter $a$) and states $3$ and $4$ belong to the second one (with persistence parameter $b$).
The probabilities $t_1$ and $t_2$ reflect transitions between these two APRWs.
We choose the transitions so that a transition from one APRW to another one finishes an ongoing zig-zag streak.
Therefore, transitions from state $1$ to state $4$ and from state $2$ to state $3$ are not allowed in our model.
These considerations lead to the transition matrix $\mathbf{T}$ in Eq.~(\ref{eq:transition_matrix}) describing the HMM in Fig.~\ref{fig:hmm1}.
The emission densities are chosen such that the sharp peaks in the distribution of spatial shifts appear in both APRWs, whereas the broader humps only occur in the first one.
In the following, we derive some analytical results for HMMs consisting of $N=4$ states with transition matrix $\mathbf{T}$ and emission densities $\lambda_i(x)$.
Note that these results can trivially be extended to HMMs with an arbitrary number of states.

The time evolution of a probability distribution $\mathbf{p}(t)=(p_1(t),p_2(t),p_3(t),p_4(t))^{\text{T}}$ consisting of the probabilities $p_i(t)$ to be in state $i$ at time $t$ is given by $\mathbf{p}(t+1)=\mathbf{T}\,\mathbf{p}(t)$.
Usually, this distribution converges to a stationary distribution as $t$ goes to infinity, $\mathbf{p}(t)\overset{t\rightarrow\infty}{\longrightarrow}\mathbf{p}^*$ with $\mathbf{p}^*=\mathbf{T}\,\mathbf{p}^*$.
For the following analytical consideration, we define the matrix $\boldsymbol{\Lambda}(x)=\text{diag}(\lambda_1(x),\lambda_2(x),\lambda_3(x),\lambda_4(x))$ and the deduced matrices $\boldsymbol{\Lambda}^+$ and $\boldsymbol{\Lambda}^-$
with $\boldsymbol{\Lambda}^+=\int_0^{\infty}\boldsymbol{\Lambda}(x)\,\text{d}x=\mathbf{I}-\boldsymbol{\Lambda}^-$ whose diagonals contain the probabilities for the occurrence of a positive or negative emission on the four states, respectively.
Furthermore, we define the diagonal matrix $\mathbf{M}=\int_{-\infty}^{\infty}x\,\boldsymbol{\Lambda}(x)\,\text{d}x$ consisting of the mean emissions on the four states.

We use the transfer matrix method, which is well established in statistical physics and has been used, e.g.,
to calculate the canonical partition function of the Ising model \cite{kramers1941_1,kramers1941_2} or the correlated output of the Preisach model of hysteresis \cite{radons2008_1,radons2008_2}.
With transfer matrix calculations, we obtain analytical expressions for the correlation of the algebraic signs of the spatial shifts, the correlation of the spatial shifts itself as well as the distribution of zig-zag streaks
in terms of the transition matrix $\mathbf{T}$, the stationary distribution $\mathbf{p}^*$, and the matrices $\boldsymbol{\Lambda}(x)$, $\boldsymbol{\Lambda}^{\pm}$, and $\mathbf{M}$ introduced above.
Here, we summarize the analytical results.
Their derivation can be found in Appendix~\ref{sec:A} or in \cite{albers2019_2}.

For the correlation function of the algebraic signs $S_t=\text{sign}(\delta x_t)$ of the spatial shifts and the correlation of the spatial shifts itself, we obtain
\begin{equation}
\begin{split}
\label{eq:jump_corr1}
C_S(\tau)&=\langle S_t\,S_{t+\tau}\rangle\\[1ex]
&=\begin{cases}1,\quad&\tau=0\\[1ex]\boldsymbol{\eta}^{\text{T}}(\boldsymbol{\Lambda}^+-\boldsymbol{\Lambda}^-)\,\mathbf{T}^{\tau}(\boldsymbol{\Lambda}^+-\boldsymbol{\Lambda}^-)\,\mathbf{p}^*,\quad&\tau>0\end{cases}
\end{split}
\end{equation}
and
\begin{equation}
\label{eq:jump_corr2}
C_{\delta x}(\tau)=\langle\delta x_t\,\delta x_{t+\tau}\rangle=\begin{cases}\langle\delta x^2\rangle,\quad&\tau=0\\[1ex]\boldsymbol{\eta}^{\text{T}}\mathbf{M}\,\mathbf{T}^{\tau}\mathbf{M}\,\mathbf{p}^*,\quad&\tau>0\end{cases},
\end{equation}
respectively.
For the distribution of zig-zag streaks, we get
\begin{equation}
\label{eq:zig-zag_dist}
p(\Delta t)=\begin{cases}2\,\boldsymbol{\eta}^{\text{T}}\boldsymbol{\Lambda}^+\mathbf{T}\boldsymbol{\Lambda}^+\left[\mathbf{T}\boldsymbol{\Lambda}^-\mathbf{T}\boldsymbol{\Lambda}^+\right]^{\Delta t/2}\mathbf{p}^*,\,&\Delta t\text{ even}\\[1ex]2\,\boldsymbol{\eta}^{\text{T}}\boldsymbol{\Lambda}^-\left[\mathbf{T}\boldsymbol{\Lambda}^-\mathbf{T}\boldsymbol{\Lambda}^+\right]^{(\Delta t+1)/2}\mathbf{p}^*,\,&\Delta t\text{ odd}\end{cases}.
\end{equation}

In order to compare these analytical findings with numerical results for the dynamics of the spatial shifts for $\mu=-0.10$,
we have to determine the parameters $a$, $b$, $t_1$, and $t_2$ of the transition matrix $\mathbf{T}$ as well as the emission densities $\lambda_i(x)$.
In Appendix~\ref{sec:B}, we demonstrate in detail how these parameters can be obtained.
For this task, we use the two exponents of the bi-exponential decay of the correlation of the algebraic signs as well as the prefactors and the second exponent of the bi-exponential decay of the distribution of zig-zag streaks.
This leads to the following system of equations:
\begin{equation}
\label{eq:equations1}
\sigma_3\approx-0.86,\,\sigma_4\approx-0.20,\,b+t_1\approx0.15,\,t_1/t_2\approx10.11,
\end{equation}
where the $\sigma_i$ are the eigenvalues of the transition matrix $\mathbf{T}$, with solution:
\begin{equation}
\label{eq:parameters1}
a\approx0.39,\quad b\approx0.001,\quad t_1\approx0.14,\quad t_2\approx0.014\,.
\end{equation}
For the emission densities, we obtain:
\begin{equation}
\begin{split}
\label{eq:emissions1}
\lambda_{1/2}(x)&\approx0.51\cdot\mathcal{N}(x|\mp0.77,0.24)\\
&\hspace{1em}+0.49\cdot\mathcal{N}(x|\mp2.16,0.026),\\[1ex]
\lambda_{3/4}(x)&\approx\mathcal{N}(x|\mp2.16,0.026)\,.
\end{split}
\end{equation}
These densities are mixtures of the Gaussian distributions $\mathcal{N}(x|\mp0.77,0.24)$ and $\mathcal{N}(x|\mp2.16,0.026)$ approximating the broader humps and the sharp peaks in the distribution of spatial shifts, respectively.
Note that one of the corresponding weights $\phi_{i1}$ or $\phi_{i2}$ of the mixture belonging to emission density $\lambda_i(x)$ as introduced in Eq.~(\ref{eq:Gaussian_mixture}) may be equal to zero.
This is the case for the broader humps, which only contribute to state $1$ and state $2$, whereas the sharp peaks must contribute to both APRWs, i.e., to the emission densities $\lambda_{1/2}(x)$ and $\lambda_{3/4}(x)$.

A comparison of our analytical results for the HMM shown in Fig.~\ref{fig:hmm1} with parameters from Eq.~(\ref{eq:parameters1}) and emission densities from Eq.~(\ref{eq:emissions1})
with the numerically determined distribution of spatial shifts in Fig.~\ref{fig:trajectories} (d), the correlation of spatial shifts, and the distribution of zig-zag streaks is shown in Figs.~\ref{fig:numerics_theory1} (b) and (c).
We obtain a remarkable agreement.
The only discrepancy in the correlation of spatial shifts for small values of $\tau$ is due to the neglect of the small humps at the margins of the distribution of spatial shifts.
Despite of their small weight, they contribute significantly to the variance of the distribution, which is identical to the correlation function for $\tau=0$.

Further insight into the soliton dynamics can be obtained by introducing a symbolic dynamics representing the soliton motion.
To do so, we replace in the sequences of spatial shifts all $\delta x_t$ with $|\delta x_t|<1.75$ by $-1$ and all $\delta x_t$ with $|\delta x_t|>1.75$ by $+1$.
In the resulting symbolic time series, we count the occurrences of so-called `plusblocks' and `minusblocks',
where a plusblock of length $\Delta t$ is defined by $\Delta t$ successive occurrences of the symbol $+1$, for instance, $-1,+1,+1,+1,-1$ for $\Delta t=3$.
Minusblocks are defined accordingly.
Figure~\ref{fig:numerics_theory1} (d) shows the absolute frequency of the occurrences of such blocks.
The distribution of the minusblocks decays mono-exponentially, whereas the distribution of plusblocks decays bi-exponentially very similar to the results for the filtered data.
These decays are well reproduced by the HMM, but interestingly, the plusblocks show even-odd oscillations, which represent higher order statistics that is not captured by our minimal model.
We want to emphasize that we only used some details of the correlation of the algebraic signs of the spatial shifts and of the distribution of zig-zag streaks to determine the parameters of the HMM.
But as a result, this model is able not only to reproduce these aspects but also other details of the statistics as well as the exponential decays of the distributions of plusblocks and minusblocks.

We want to emphasize that the derived HMM for the dynamics of the spatial shifts for $\mu=-0.10$ in Fig.~\ref{fig:hmm1} is a special case of the HMM in Fig.~\ref{fig:hmm}
with the weights $\phi_{31}$ and $\phi_{41}$ of the broader humps in states $3$ and $4$ equal to zero.

\section{\label{sec:V}Connecting Anti-Persistent Random Walks and Hidden Markov Models}

\begin{figure}
\includegraphics[width=\linewidth]{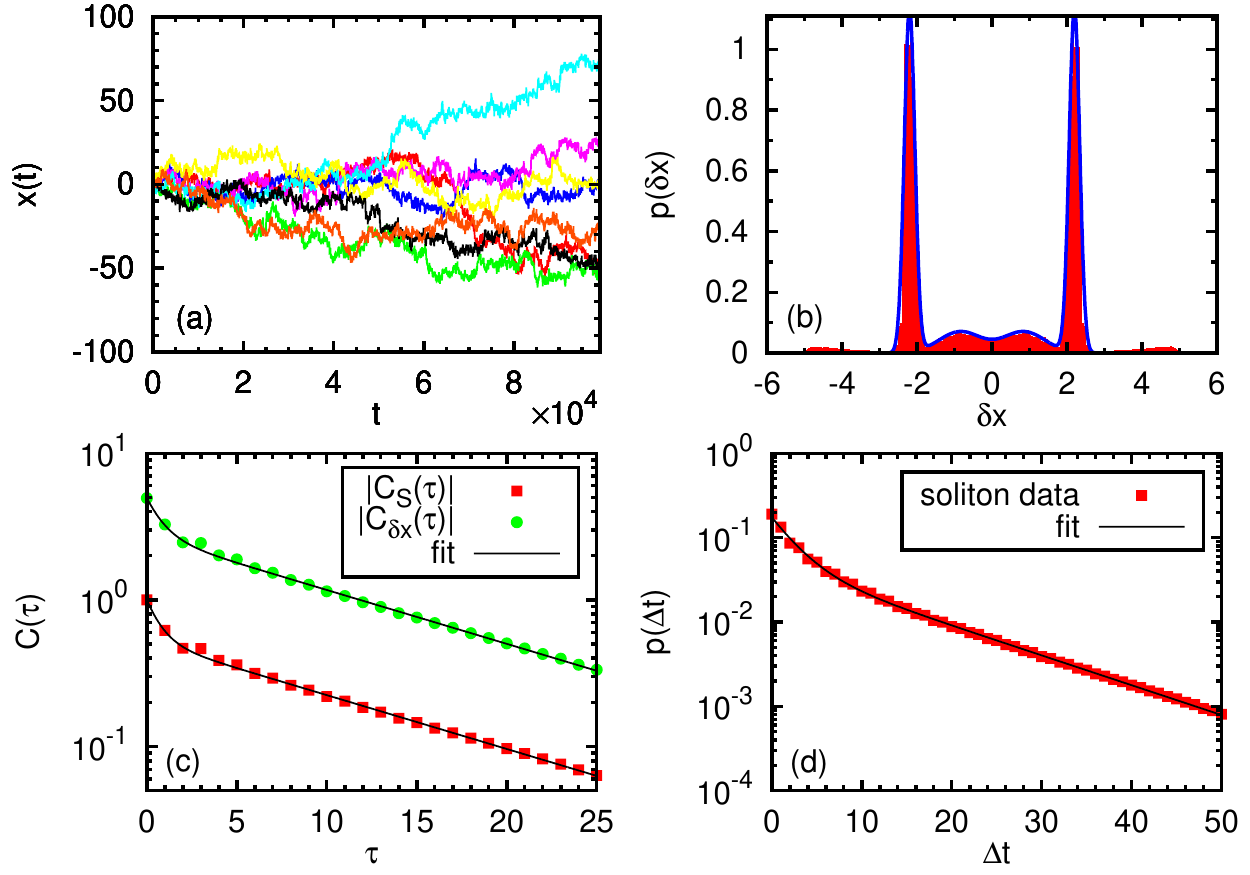}
\caption{\label{fig:statistics2}(color online)
(a) Eight independent realizations $x(t)$ of the soliton motion for $\mu=-0.13$ on a large time scale.
(b) Numerically determined distribution $p(\delta x)$ of spatial shifts for $\mu=-0.13$ (red histogram) compared with a weighted sum of four Gaussian distributions (blue line),
$0.1\cdot\mathcal{N}(x|\pm0.85,0.32)+0.4\cdot\mathcal{N}(x|\pm2.2,0.02)$, similar to the distribution shown in Fig.~\ref{fig:trajectories}(d) for $\mu=-0.10$.
The Gaussian mixture is reproduced by the Hidden Markov Model shown in Fig.~\ref{fig:hmm1} with parameters and emission densities from Eq.~(\ref{eq:parameters2}) and Eq.~(\ref{eq:emissions2}), respectively.
Again, the small humps on the margin of the distribution are neglected by the Hidden Markov Model.
(c) Absolute value of the numerically determined correlations $C_{\delta x}(\tau)$ of spatial shifts $\delta x_t$ (green circles) and $C_S(\tau)$ of the algebraic signs $S_t$ of the spatial shifts (red squares) for $\mu=-0.13$.
Again as in Fig.~\ref{fig:statistics1}(a) for $\mu=-0.10$, the black lines are corresponding bi-exponential fits, $2.24\cdot\exp(-1.12\cdot\tau)+2.72\cdot\exp(-0.085\cdot\tau)$ and $0.48\cdot\exp(-1.27\cdot\tau)+0.52\cdot\exp(-0.085\cdot\tau)$, respectively.
(d) Numerically determined distribution $p(\Delta t)$ of zig-zag streaks for $\mu=-0.13$.
The black line is a bi-exponential fit, $0.13\cdot\exp(-0.39\cdot\Delta t)+0.046\cdot\exp(-0.081\cdot\Delta t)$.}
\end{figure}

An interesting observation from the results of the previous section is that the persistence parameter $b$ of the Hidden Markov Model (HMM) for the dynamics of the spatial shifts for $\mu=-0.10$ in Fig.~\ref{fig:hmm1}
is nearly identical to the persistence parameter $a$ of the Anti-Persistent Random Walk (APRW) that describes the soliton motion for $\mu=-0.18$ and whose generator is the $2$-state Markov model in Fig.~\ref{fig:aprw}.
This leads to the assumption that the $2$-state Markov model is a special case of the more general HMM with only states $3$ and $4$ being present.
But how can this be?
How can a continuous change of the bifurcation parameter $\mu$ of the CGL transform a $2$-state Markov model into a more general HMM and vice versa?
A priori, several scenarios are conceivable.
For instance, the transition probability $t_1$ from states $3$ and $4$ of the HMM to states $1$ and $2$ could go to zero so that only states $3$ and $4$ survive asymptotically.
Another possibility is that the persistence parameters $a$ and $b$ of the HMM as well as the emission densities $\lambda_{1/2}(x)$ and $\lambda_{3/4}(x)$ become identical.
In order to understand the transition from the $2$-state Markov model for $\mu=-0.18$ to the HMM for $\mu=-0.10$ in more detail,
we investigate the soliton dynamics for an intermediate value of the bifurcation parameter, $\mu=-0.13$, where according to Fig.~\ref{fig:trajectories} (c), the broader humps in the distribution of spatial shifts begin to appear.
The numerical results for $\mu=-0.13$ are shown in Fig.~\ref{fig:statistics2}.

\begin{figure}
\includegraphics[width=\linewidth]{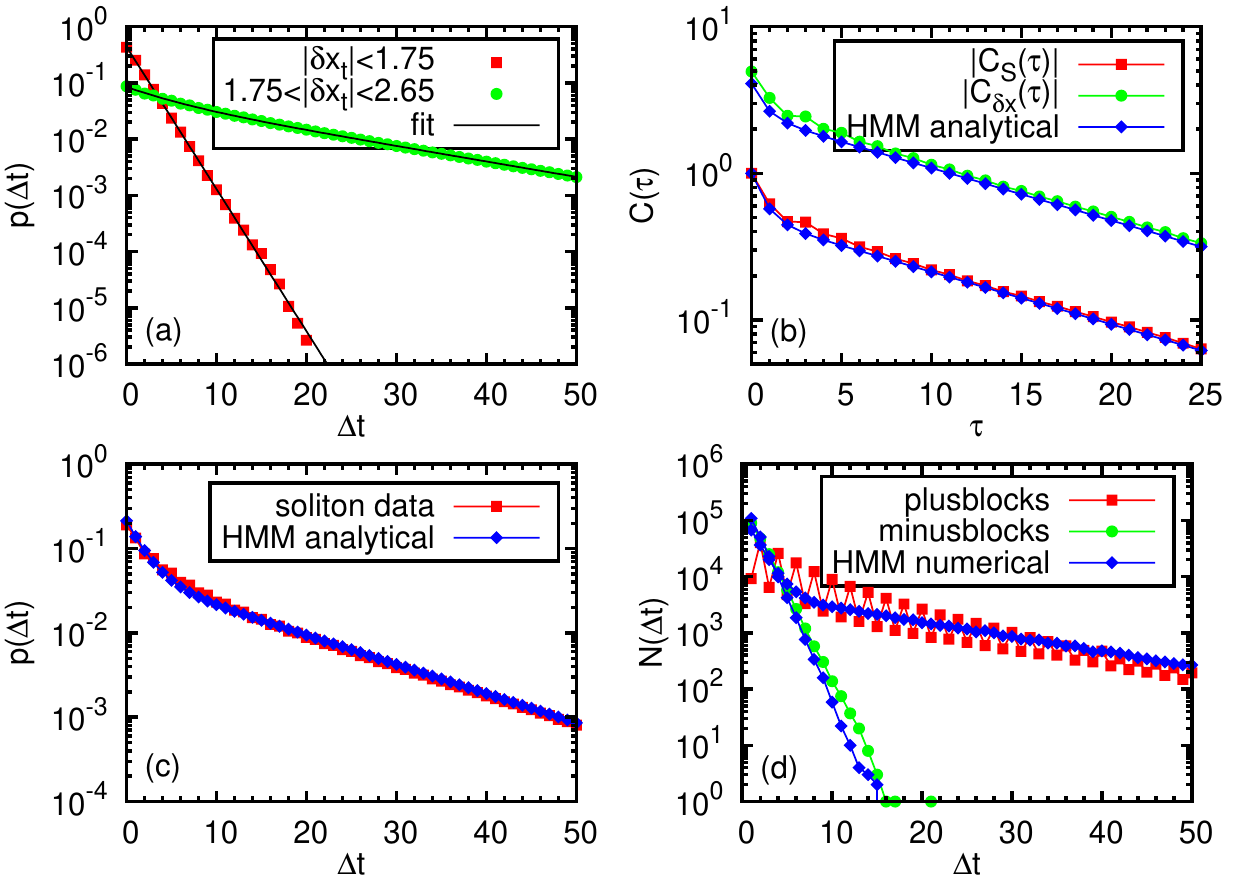}
\caption{\label{fig:numerics_theory2}(color online)
(a) Distribution $p(\Delta t)$ of zig-zag streaks for two filtered sequences of spatial shifts,
where the first one contains all $\delta x_t$ with $|\delta x_t|<1.75$ (red squares) and the second one contains all $\delta x_t$ with $1.75<|\delta x_t|<2.65$ (green circles).
The black lines are corresponding fits, where the first one is mono-exponential, $0.44\cdot\exp(-0.58\cdot\Delta t)$, and the second one is bi-exponential, $0.035\cdot\exp(-0.22\cdot\Delta t)+0.050\cdot\exp(-0.063\cdot\Delta t)$,
similar to the situation shown in Fig.~\ref{fig:numerics_theory1}(a) for $\mu=-0.10$.
(b) Comparison of the numerically determined correlations $C_{\delta x}(\tau)$ of spatial shifts $\delta x_t$ (green circles) and $C_S(\tau)$ of the algebraic signs $S_t$ of the spatial shifts (red squares) for $\mu=-0.13$
with the corresponding analytical results (blue diamonds) for the HMM shown in Fig.~\ref{fig:hmm1} with parameters and emission densities from Eq.~(\ref{eq:parameters2}) and Eq.~(\ref{eq:emissions2}), respectively.
(c) Numerically determined distribution $p(\Delta t)$ of zig-zag streaks for $\mu=-0.13$ compared with the corresponding analytical result from the HMM.
(d) Absolute frequency for the occurrence of plusblocks and minusblocks numerically determined for $\mu=-0.13$ and compared with corresponding numerical results obtained from the HMM.}
\end{figure}

We observe that the correlation of the algebraic signs of the spatial shifts, the correlation of the spatial shifts itself as well as the distribution of zig-zag streaks decay bi-exponentially in a similar way to the case $\mu=-0.10$.
Moreover, in Fig.~\ref{fig:numerics_theory2} (a), we can see that also the distribution of zig-zag streaks for appropriate filtered sequences of spatial shifts
belonging to the broader humps or the sharper peaks in the distribution of spatial shifts show a mono-exponential or a bi-exponential decrease, respectively.
Therefore, we can conclude that the HMM shown in Fig.~\ref{fig:hmm1} should also describe the dynamics of the spatial shifts for $\mu=-0.13$.
For the determination of the parameters of the HMM, we can follow exactly the same procedure as in the case $\mu=-0.10$, which is described in detail in Appendix~\ref{sec:B}.
Again, we use the bi-exponential fits from the correlation of algebraic signs of spatial shifts and the distribution of zig-zag streaks in order to obtain the following system of equations:
\begin{equation}
\label{eq:equations2}
\sigma_3\approx-0.92,\,\sigma_4\approx-0.28,\,b+t_1\approx0.078,\,t_1/t_2\approx0.67
\end{equation}
with solution:
\begin{equation}
\label{eq:parameters2}
a\approx0.31,\quad b\approx0.012,\quad t_1\approx0.065,\quad t_2\approx0.098\,.
\end{equation}
For the emission densities, we get:
\begin{equation}
\begin{split}
\label{eq:emissions2}
\lambda_{1/2}(x)&\approx0.5\cdot\mathcal{N}(x|\mp0.85,0.32)\\
&\hspace{1em}+0.5\cdot\mathcal{N}(x|\mp2.2,0.02),\\[1ex]
\lambda_{3/4}(x)&\approx\mathcal{N}(x|\mp2.2,0.02)\,.
\end{split}
\end{equation}

Figures~\ref{fig:numerics_theory2} (b), (c), and (d) show a comparison of the correlation functions, the distribution of zig-zag streaks and the statistics of plus- and minusblocks with the HMM from Fig.~\ref{fig:hmm1}
with parameters from Eq.~(\ref{eq:parameters2}) and emission densities from Eq.~(\ref{eq:emissions2}).
We can see an excellent agreement, even better than in the case $\mu=-0.10$.

With the transition probabilities $t_1$ and $t_2$ and the stationary distribution $\mathbf{p}^*=(p_1^*,p_2^*,p_3^*,p_4^*)^{\text{T}}$ in Eq.~(\ref{eq:stationary_distribution}) of Appendix~\ref{sec:B},
we can calculate the weights $w_1=p_1^*+p_2^*$ and $w_2=p_3^*+p_4^*$ of two APRWs,
where the first one is generated by the dynamics of the spatial shifts represented by the states $1$ and $2$ of the HMM in Fig.~\ref{fig:hmm1} and the second one is generated by the states $3$ and $4$.
For $\mu=-0.10$, we obtain $w_1=0.91$ and $w_2=0.09$, and for $\mu=-0.13$, we get $w_1=0.40$ and $w_2=0.60$.
The important point to realize is that the weight $w_1$ of the first APRW decreases for a decreasing value of $\mu$.
When the value of the bifurcation parameter $\mu$ becomes even smaller, this weight goes to zero and only the second APRW generated by the states $3$ and $4$ of the original HMM remains.
This APRW can be identified with the one that was found and studied in detail in a previous publication \cite{albers2019_1} for $\mu=-0.18$.

\begin{figure}
\includegraphics[width=\linewidth]{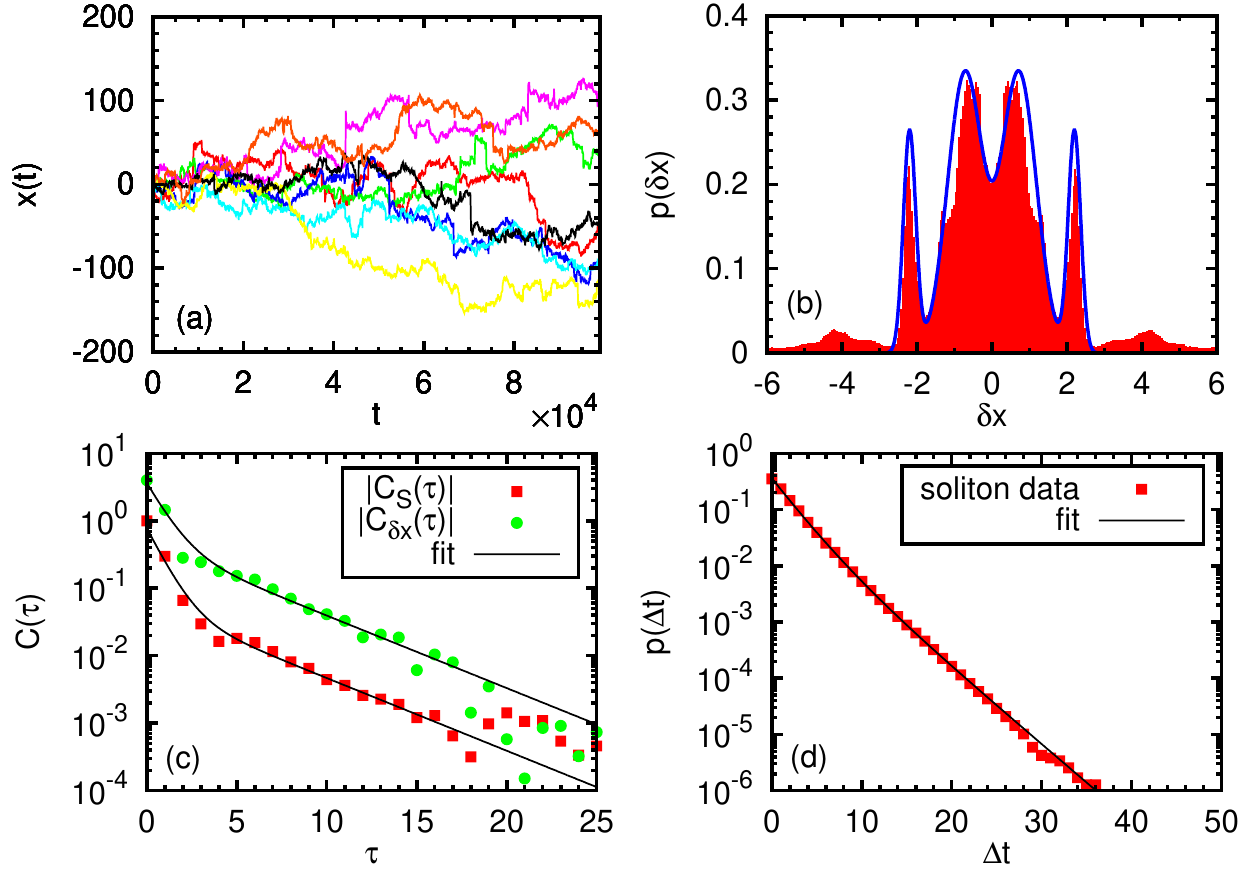}
\caption{\label{fig:statistics3}(color online)
The same statistical analysis as in Fig.~\ref{fig:statistics2} but now for a larger value of the bifurcation parameter, $\mu=-0.03$.
In contrast to the cases $\mu=-0.13$ and $\mu=-0.10$,
the broader humps in the distribution $p(\delta x)$ of spatial shifts in Figure (b) have a much larger weight leading to significant changes of the emission densities of the associated HMM shown in Fig.~\ref{fig:hmm2}.
This model with parameters from Eq.~(\ref{eq:parameters3}) and emission densities from Eq.~(\ref{eq:emissions3}) leads to a Gaussian mixture of the output density (blue line in Figure (b)),
$0.39\cdot\mathcal{N}(x|\pm0.72,0.22)+0.11\cdot\mathcal{N}(x|\pm2.2,0.028)$, that approximates the distribution of spatial shifts quite well.
The numerically determined correlations $C_{\delta x}(\tau)$ of spatial shifts $\delta x_t$ (green circles) and $C_S(\tau)$ of the algebraic signs $S_t$ of the spatial shifts (red squares) in Figure (c) decay bi-exponentially again.
The corresponding fits, $3.17\cdot\exp(-1.13\cdot\tau)+0.47\cdot\exp(-0.25\cdot\tau)$ and $0.72\cdot\exp(-1.24\cdot\tau)+0.056\cdot\exp(-0.25\cdot\tau)$, respectively, are drawn by black lines.
Also the distribution of zig-zag streaks in Figure (d) decays bi-exponentially.
The associated fit is given by $0.28\cdot\exp(-0.51\cdot\Delta t)+0.083\cdot\exp(-0.31\cdot\Delta t)$.}
\end{figure}

In a next step, we want to investigate what happens to the HMM if we change the bifurcation parameter $\mu$ in the opposite direction.
It is natural to assume that if the value of the bifurcation parameter $\mu$ becomes larger than $-0.10$, the weight of the first APRW further increases and the weight of the second APRW further decreases.
In order to check this hypothesis, we investigate the soliton dynamics for $\mu=-0.03$.
The corresponding numerical results are shown in Fig.~\ref{fig:statistics3}.

\begin{figure}
\includegraphics[width=\linewidth]{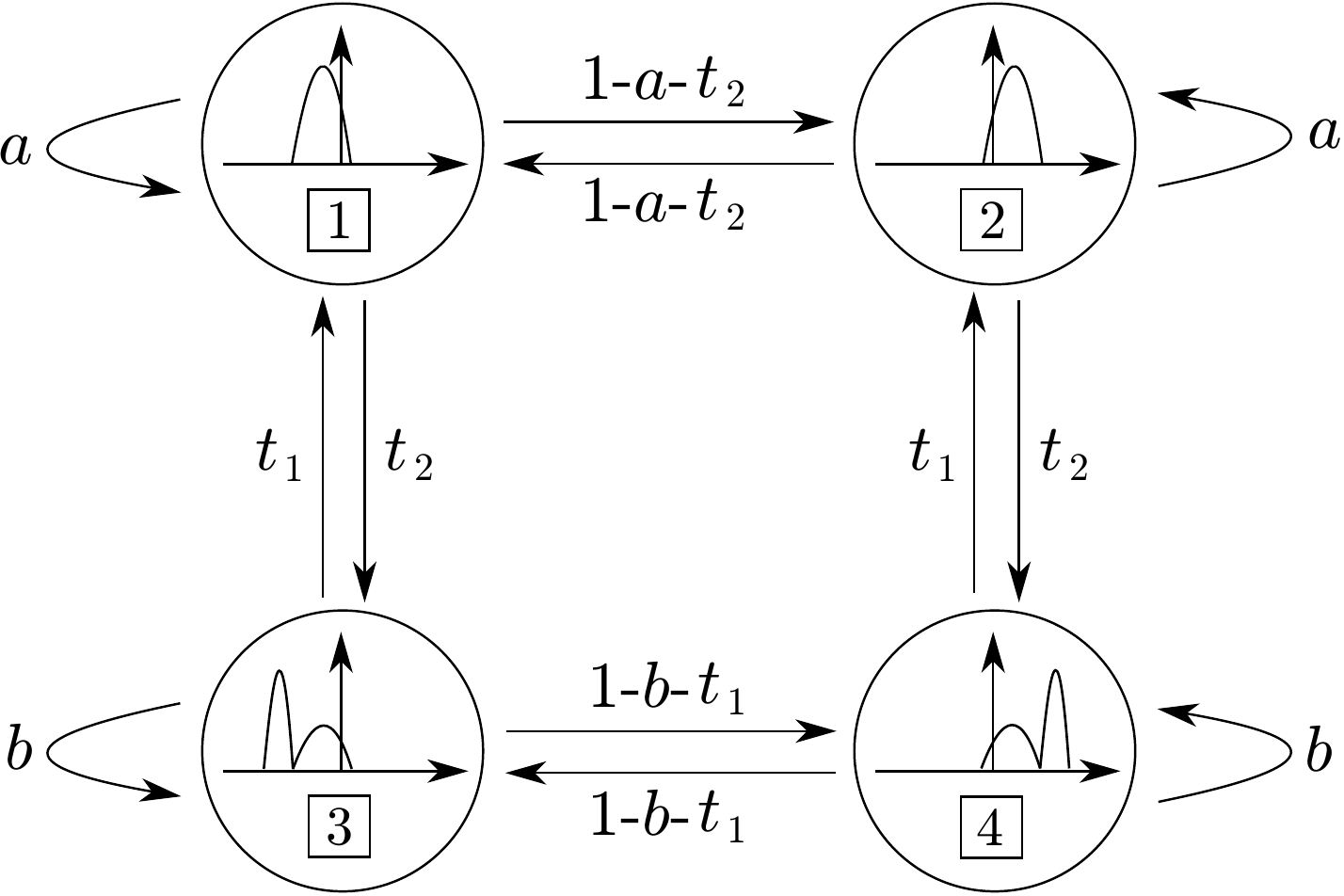}
\caption{\label{fig:hmm2}
Schematic illustration of a Hidden Markov Model for the dynamics of the spatial shifts $\delta x_t$ for $\mu=-0.03$ whose distribution is shown in Fig.~\ref{fig:statistics3} (b).}
\end{figure}

Again, we find bi-exponentially decaying correlations and distributions of zig-zag streaks.
An analysis of the bi-exponential fit of the distribution of zig-zag streaks according to the evaluation in Appendix~\ref{sec:B} for $\mu=-0.10$ gives the weights $w_1$ and $w_2$ of the two APRWs for $\mu=-0.03$.
We obtain $w_1=0.69$ and $w_2=0.31$, thus violating our hypothesis.
Interestingly, based on the HMM shown in Fig.~\ref{fig:hmm1}, these weights are incompatible with the weight of the broader humps in the distribution of spatial shifts in Fig.~\ref{fig:statistics3} (b).
How can the broader humps capture roughly $80$ percent of the weight in the distribution of spatial shifts when they only appear in the first APRW that captures roughly $70$ percent of the HMM?
The answer is that now the broader humps contribute to both APRW, whereas the sharper peaks contribute to only one of them.
This leads to the modified HMM shown in Fig.~\ref{fig:hmm2}.

This new HMM changes the procedure of the determination of its parameters in one point compared to the procedure described in Appendix~\ref{sec:B}.
The second exponential decrease of the distribution of zig-zag streaks is now additionally influenced by the fact that, for instance, a positive emission on state $4$ is only obtained with probability $\lambda_4^+<1$.
This means that the second exponential decrease of the distribution of zig-zag streaks should be proportional to $((1-b-t_1)\lambda_4^+)^{\Delta t}$.
The rest of the procedure remains unchanged leading to the following system of equations:
\begin{equation}
\label{eq:equations3}
\sigma_3\approx-0.78,\,\sigma_4\approx-0.29,\,b+t_1\approx0.26,\,t_1/t_2\approx2.23
\end{equation}
with solution:
\begin{equation}
\label{eq:parameters3}
a\approx0.27,\quad b\approx0.021,\quad t_1\approx0.24,\quad t_2\approx0.11\,.
\end{equation}
The emission densities are now given by:
\begin{equation}
\begin{split}
\label{eq:emissions3}
\lambda_{1/2}(x)&\approx\mathcal{N}(x|\mp0.72,0.22),\\[1ex]
\lambda_{3/4}(x)&\approx0.29\cdot\mathcal{N}(x|\mp0.72,0.22)\\
&\hspace{1em}+0.71\cdot\mathcal{N}(x|\mp2.2,0.028)\,.
\end{split}
\end{equation}

\begin{figure}
\includegraphics[width=\linewidth]{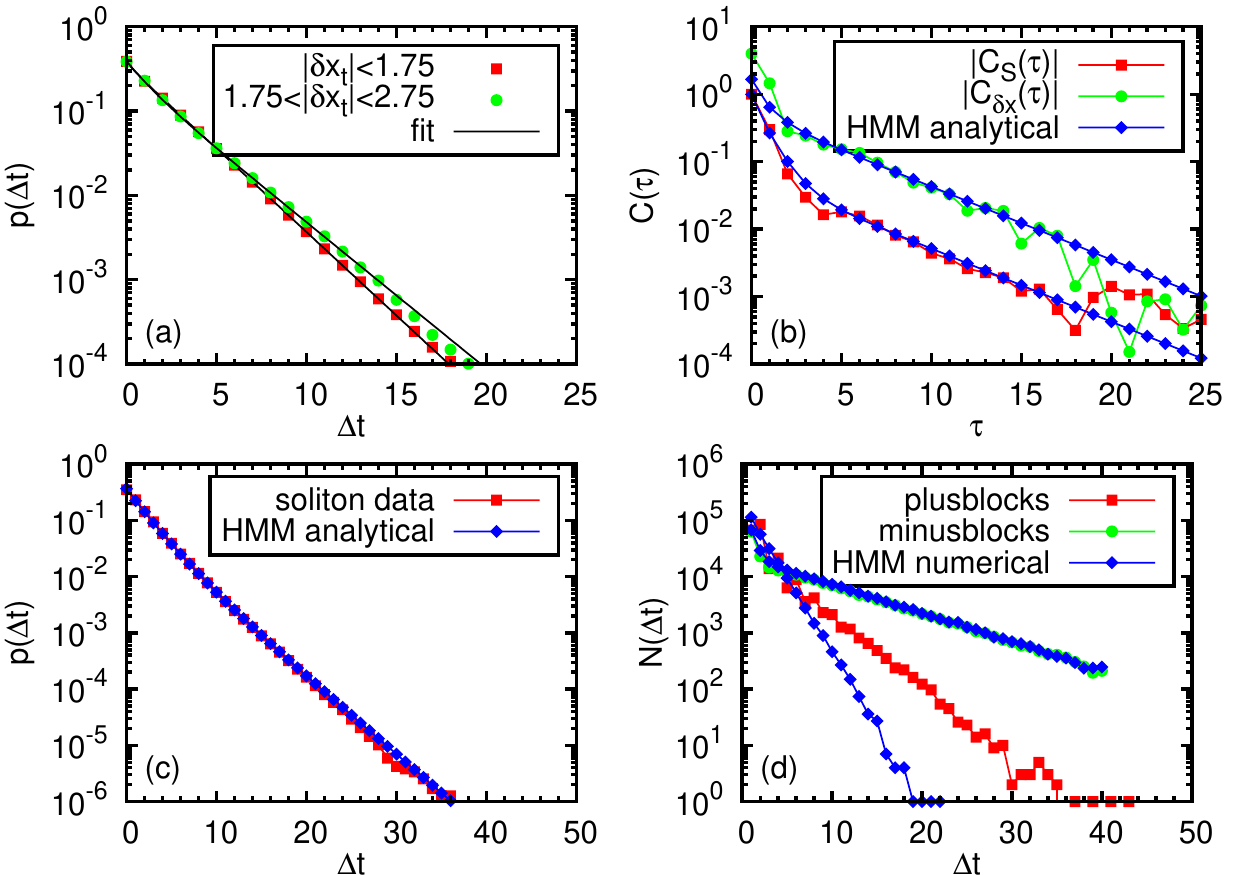}
\caption{\label{fig:numerics_theory3}(color online)
The same statistical quantities as in Fig.~\ref{fig:numerics_theory2} numerically determined for the larger bifurcation parameter $\mu=-0.03$
and compared with corresponding analytical results (blue diamonds) for the HMM shown in Fig.~\ref{fig:hmm2} with parameters from Eq.~(\ref{eq:parameters3}) and emission densities from Eq.~(\ref{eq:emissions3}).
In Figure (a), the distribution $p(\Delta t)$ of zig-zag streaks is shown for two filtered sequences of spatial shifts,
where the first one corresponds to all $\delta x_t$ belonging to the broader humps (red squares) and the second one contains all $\delta x_t$ belonging to the sharp peaks (green circles).
The black lines are bi-exponential fits, $0.037\cdot\exp(-2.32\cdot\Delta t)+0.35\cdot\exp(-0.46\cdot\Delta t)$ and $0.13\cdot\exp(-0.96\cdot\Delta t)+0.26\cdot\exp(-0.40\cdot\Delta t)$, respectively.
In Figures (b), (c), and (d), the correlations $C_{\delta x}(\tau)$ of spatial shifts $\delta x_t$ and $C_S(\tau)$ of the algebraic signs $S_t$ of the spatial shifts,
the distribution of zig-zag streaks, and the absolute frequencies for the occurrence of plusblocks and minusblocks are compared with analytical results for the corresponding HMM.}
\end{figure}

Figure~\ref{fig:numerics_theory3} compares all numerically determined statistics of the soliton motion for $\mu=-0.03$ with the HMM shown in Fig.~\ref{fig:hmm2}
with parameters from Eq.~(\ref{eq:parameters3}) and emission densities from Eq.~(\ref{eq:emissions3}).
Again, we find a very good agreement.
However, whereas the bi-exponential decrease of the distribution of minusblocks is very well reproduced, our model fails in reproducing the second exponential decay of the distribution of plusblocks.
A possible reason is the neglect of the small humps at the margins of the distribution of spatial shifts shown in Fig.~\ref{fig:statistics3} (b), which are not incorporated in our minimal model.

Note that similar to the HMMs found for $\mu=-0.13$ and $\mu=-0.10$, also the HMM for $\mu=-0.03$ in Fig.~\ref{fig:hmm2} is a special case of the HMM shown in Fig.~\ref{fig:hmm}
with the weights $\phi_{12}$ and $\phi_{22}$ of the sharp peaks in states $1$ and $2$ equal to zero.

In Table~\ref{tab:parameters} and \ref{tab:emissions}, all parameters of the HMMs that we determined for the different values of the bifurcation parameter $\mu$ are summarized.
From Tab.~\ref{tab:emissions}, we can see that for the cases $\mu=-0.13$ and $\mu=-0.10$, the emission densities are nearly identical.
Moreover, the emission densities on states $3$ and $4$ almost coincide with the emission densities on states $3$ and $4$ for the case $\mu=-0.18$ (except for very small changes of the mean and the variance of the Gaussian distributions).
Because of that and the fact that the weight $w_1$ of the first APRW that is generated by states $1$ and $2$ is smaller for $\mu=-0.13$ than for $\mu=-0.10$,
it is natural to assume that this weight further decreases for further decreasing bifurcation parameter $\mu$.
This leads to the conclusion that this weight goes to zero, and for $\mu=-0.18$, only the second APRW generated by states $3$ and $4$, which was already found in \cite{albers2019_1}, remains.
Therefore, the transition from the single APRW for $\mu=-0.18$ generated by a simple $2$-state Markov process to the more complex HMM for larger values of $\mu$ occurs by the appearance of another APRW with increasing weight.
For the largest bifurcation parameter in our investigation, $\mu=-0.03$, the weight $w_1$ is also smaller than for the case $\mu=-0.10$, but here also the emissions on the four states change significantly.

\begin{table}
\begin{tabular}{|c||c|c|c|c|c|c|}\hline
$\mu$&$a$&$b$&$t_1$&$t_2$&$w_1$&$w_2$\\ \hline\hline
-0.18&--&0.001&0&--&0&1\\ \hline
-0.13&0.31&0.012&0.065&0.098&0.40&0.60\\ \hline
-0.10&0.39&0.001&0.14&0.014&0.91&0.09\\ \hline
-0.03&0.27&0.021&0.24&0.11&0.69&0.31\\ \hline
\end{tabular}
\caption{\label{tab:parameters}Summary of the determined structural parameters of the Hidden Markov Models
for the dynamics of the spatial shifts of the soliton motion for different values of the bifurcation parameter $\mu$ of the complex Ginzburg-Landau equation.}
\end{table}

\begin{table}
\begin{tabular}{|c||c|c|c|c|}\hline
$\mu$&$\phi_{11}$&$\phi_{12}$&$\phi_{31}$&$\phi_{32}$\\ \hline\hline
-0.18&--&--&0&1\\ \hline
-0.13&0.50&0.50&0&1\\ \hline
-0.10&0.51&0.49&0&1\\ \hline
-0.03&1&0&0.29&0.71\\ \hline
\end{tabular}
\caption{\label{tab:emissions}Summary of the determined weights $\phi_{ij}$, defined in Eq.~(\ref{eq:Gaussian_mixture}) and Eq.~(\ref{eq:emissions}), of the broader humps ($j=1$) and the sharp peaks ($j=2$)
for the emission densities $\lambda_i(x)$ on state $i$ of the Hidden Markov Models shown in Figs.~\ref{fig:hmm1} and \ref{fig:hmm2} for different values of the bifurcation parameter $\mu$.
Due to symmetry of the emissions, $\phi_{1j}=\phi_{2j}$ represents the first Anti-Persistent Random Walk and $\phi_{3j}=\phi_{4j}$ the second one.}
\end{table}

\section{\label{sec:VI}Discussion}

In previous publications \cite{albers2019_1,albers2019_2}, we investigated the soliton dynamics governed by the complex cubic-quintic Ginzburg-Landau equation for two specific values of the bifurcation parameter $\mu$.
For $\mu=-0.18$, we found that the soliton motion can be well described by an Anti-Persistent Random Walk (APRW) whose sequences of spatial shifts are generated by the simple $2$-state Markov process that is shown in Fig.~\ref{fig:aprw}.
For $\mu=-0.10$, however, we realized that the latter has to be replaced by the more complex Hidden Markov Model (HMM) in Fig.~\ref{fig:hmm1},
where states $1$ and $2$ as well as states $3$ and $4$ can be interpreted as generators of two APRWs with transitions between them captured by the probabilities $t_1$ and $t_2$.
In accordance with these findings, correlations in the soliton dynamics decrease mono-exponentially for $\mu=-0.18$, whereas for $\mu=-0.10$, bi-exponential decays of correlations were found.
These previous studies posed the question of how a continuous change of the bifurcation parameter $\mu$ of the underlying nonlinear partial differential equation
transforms a simple $2$-state Markov process into a more complex HMM and vice versa.

At least, two possible scenarios are obvious: one of the two involved APRWs of the HMM vanishes in some way or both APRWs become identical.
In the first case, if the ratio of the transition probabilities $t_1$ and $t_2$, which solely determines the stationary distribution $\mathbf{p}^*$ according to Eq.~(\ref{eq:stationary_distribution}), decreases,
the weight $w_1=p_1^*+p_2^*$ of the first APRW generated by states $1$ and $2$ of the HMM also becomes smaller.
In the limit $t_1\rightarrow0$, only states $3$ and $4$ of the HMM generating the second APRW asymptotically remain because the first APRW behaves as a repeller.
Interestingly, in this case, the two eigenvalues $\sigma_3$ and $\sigma_4$ in Eq.~(\ref{eq:eigenvalues}) of the transition matrix $\mathbf{T}$ in Eq.~(\ref{eq:transition_matrix}) do not coincide.
Nevertheless, the correlation functions decay mono-exponentially due to a vanishing contribution from the eigenvalue $\sigma_4$ because the corresponding dyadic product of right and left eigenvectors
multiplied with the vectors and matrices appearing in Eq.~(\ref{eq:jump_corr1}) and Eq.~(\ref{eq:jump_corr2}) gives zero for $t_1\rightarrow0$.
Therefore, the eigenvalue $\sigma_3$ determines the mono-exponential decay of the correlation functions.
In the second scenario, because of the vanishing broader humps in the distribution of spatial shifts for decreasing values of the bifurcation parameter $\mu$, see Fig.~\ref{fig:trajectories} (c),
one could assume that they also disappear in the emission densities of the states $1$ and $2$.
If additionally the structural parameters of the HMM change such that $a=b$ and $t_1=t_2$, both APRWs are identical.
Also in this second case, the eigenvalues $\sigma_3$ and $\sigma_4$ do not coincide.
Nevertheless, similar to the first case and because of the symmetry of the emissions, i.e., $\lambda_1(x)=\lambda_3(x)=\lambda_2(-x)=\lambda_4(-x)$, only one of the eigenvalues contributes to the correlation functions.
However, in this second scenario, it is the eigenvalue $\sigma_4$ that determines the mono-exponential decay of the correlation functions.

With our investigation of the soliton dynamics for $\mu=-0.13$ in this article, where we found that the transition probability $t_1$ becomes smaller compared to the case $\mu=-0.10$,
whereas the emission densities on the four states remain unchanged, we showed that the first scenario is taking place.
The weight $w_1$ of the first APRW becomes smaller when the bifurcation parameter $\mu$ is decreased from $-0.10$ to $-0.13$ and eventually goes to zero
explaining why the soliton dynamics for $\mu=-0.18$ can be well described by a single APRW as it was found in a previous article \cite{albers2019_1}.
Moreover, we also studied how the HMM changes if the parameter $\mu$ is varied in the opposite direction, i.e., if it becomes larger than $-0.10$.
For the case $\mu=-0.03$, we found that beside the changes in the structural parameters of the underlying Markov process, also the emission densities of the four states change significantly.
Because of the increased transition probabilities $t_1$ and $t_2$ for this case,
the anti-persistent nature of states $1$ and $2$ as well as states $3$ and $4$ captured by the probabilities $1-a-t_2$ and $1-b-t_1$, respectively, becomes weaker.
Therefore, it seems that for larger values of the bifurcation parameter $\mu$, the concept of two coupled APRWs starts to disintegrate.
This view is supported also by our observation that near $\mu=-0.03$ the waiting times start to be replaced by a new sub-process without jumps, namely one with symmetric explosions \cite{descalzi2010}.
This obviously does not affect our treatment of the diffusive properties via Hidden Markov Models since only the character of the waiting times is refined.
An explicit inclusion of such non-transporting states with symmetric explosions is possible if desired, but would require for its description a more complex HMM with additional states.

\section{\label{sec:VII}Summary}

In summary, we have shown that the infinite-dimensional dynamics of the cubic-quintic complex Ginzburg-Landau equation can be reduced to a one-dimensional dynamics in a physically meaningful way
by considering the center of mass of the solitons in dependence on time, i.e., the soliton motion.
The incremental process of the latter is well described by a finite state hidden Markov process, which is a minimal model able to reproduce the statistics under consideration.
This model leads to a superposition of two Anti-Persistent Random Walks for the soliton motion,
where the occurrence of the second Anti-Persistent Random Walk for larger values of the bifurcation parameter $\mu$ of the Ginzburg-Landau equation might belong to the appearance of a further attractor in the system
demonstrating that this statistical analysis enriches the understanding of the underlying nonlinear dynamics.
Furthermore, we have analyzed the Hidden Markov Model for several values of the bifurcation parameter $\mu$.
This analysis revealed the transition from the single Anti-Persistent Random Walk for smaller values of $\mu$ to the more general Hidden Markov Model for larger values of $\mu$
by the appearance of another Anti-Persistent Random Walk with continuously increasing weight during the transition.
In addition, such a surprising relation between partial differential equations and Hidden Markov Models connects soliton diffusion with the theory of complexity of dynamical systems as pioneered for low-dimensional systems in \cite{grassberger1986,crutchfield1989}
and in this spirit with recent results for entropy rates of Hidden Markov Models, see e.g. \cite{marcus2011,breitner2017}.
Note that the inclusion of (physically relevant) higher-order terms in the Ginzburg-Landau equation can destroy the left-right symmetry thus leading to drifting motion \cite{gurevich2019}.
One expects that in these more general systems, chaotic diffusion and drift can occur together.
Furthermore, because of the general relation between partial differential equations and time-delayed systems \cite{yanchuk2017}, explosions were also found in the latter \cite{schelte2019}.
We expect that our approach can be extended to these systems.

\appendix

\section{\label{sec:A}Derivation of the analytical results for the Hidden Markov Model}

\subsection{Derivation of Eq.~(\ref{eq:jump_corr2})}

In the following, we use the transfer matrix method to derive an analytical expression for the correlation of the spatial shifts for the Hidden Markov Model (HMM) shown in the main text in Fig.~\ref{fig:hmm}.
We start with the definition of the correlation function (covariance function) of the spatial shifts $\delta x_t$ using the property $\langle\delta x_t\rangle=0$
and the probability density $p(\delta x,t;\delta x',t+\tau)$ to find a spatial shift $\delta x$ at discrete time $t$ and a spatial shift $\delta x'$ at discrete time $t+\tau$,
\begin{equation}
\begin{split}
\label{eq:A1}
C_{\delta x}(\tau)&=\langle\delta x_t\,\delta x_{t+\tau}\rangle\\
&=\int_{-\infty}^{\infty}\int_{-\infty}^{\infty}\delta x\,\delta x'\,p(\delta x,t;\delta x',t+\tau)\,\text{d}(\delta x)\,\text{d}(\delta x')\,.
\end{split}
\end{equation}
As a next step, we introduce the probability (density) \footnote{$p(\delta x,i,t;\delta x',j,t+\tau)$ is a probability density with respect to $\delta x$ and $\delta x'$ and a probability with respect to $i$ and $j$}
$p(\delta x,i,t;\delta x',j,t+\tau)$ of being in state $i$ of the HMM at time $t$ and in state $j$ at time $t+\tau$ with corresponding emissions $\delta x$ and $\delta x'$, respectively,
and the corresponding conditional probability (density) $p(\delta x',j,t+\tau|\delta x,i,t)$ and use it to calculate the probability density appearing in Eq.~(\ref{eq:A1}),
\begin{equation}
\begin{split}
\label{eq:A2}
p(\delta x,t;\delta x',t+\tau)=\sum\limits_{i=1}^4\sum\limits_{j=1}^4p(\delta x,i,t;\delta x',j,t+\tau)&\\
=\sum\limits_{i=1}^4\sum\limits_{j=1}^4p(\delta x',j,t+\tau|\delta x,i,t)\,p(\delta x,i,t)&\,.
\end{split}
\end{equation}
The probability (density) $p(\delta x,i,t)$ to be in state $i$ at time $t$ with emission $\delta x$ (distributed according to $\lambda_i(\delta x)$)
can be written for large times $t$, where the process converges to the stationary distribution $\mathbf{p}^*$, as
\begin{equation}
\label{eq:A3}
p(\delta x,i,t)=\lambda_i(\delta x)\,\hat{\mathbf{e}}_i^{\text{T}}\,\mathbf{p}^*\,,
\end{equation}
where $\hat{\mathbf{e}}_i$ is the unit vector in the $i^{\text{th}}$ direction, i.e., $(\hat{\mathbf{e}}_i)_j=\delta_{ij}$.
By using the transition matrix (transfer matrix) $\mathbf{T}$ of the HMM in Eq.~(\ref{eq:transition_matrix}) of the main text, the conditional probability (density) $p(\delta x',j,t+\tau|\delta x,i,t)$ can be written as
\begin{equation}
\label{eq:A4}
p(\delta x',j,t+\tau|\delta x,i,t)=\lambda_j(\delta x')\,\hat{\mathbf{e}}_j^{\text{T}}\,\mathbf{T}^{\tau}\,\hat{\mathbf{e}}_i\,.
\end{equation}
Here, we used that the emission $\delta x'$ at time $t+\tau$ does not depend on the emission $\delta x$ at time $t$ for $\tau>0$ implying that this derivation is correct for $\tau>0$.
For $\tau=0$, the correlation function of the spatial shifts is equal to the second moment of the spatial shifts.
By multiplying Eq.~(\ref{eq:A4}) with Eq.~(\ref{eq:A3}), we get
\begin{equation}
\begin{split}
\label{eq:A5}
p(\delta x,i,t&;\delta x',j,t+\tau)\\
&=\lambda_j(\delta x')\,\hat{\mathbf{e}}_j^{\text{T}}\,\mathbf{T}^{\tau}\,\hat{\mathbf{e}}_i\,\lambda_i(\delta x)\,\hat{\mathbf{e}}_i^{\text{T}}\,\mathbf{p}^*\\
&=\lambda_j(\delta x')\,\hat{\mathbf{e}}_j^{\text{T}}\,\mathbf{T}^{\tau}\,\lambda_i(\delta x)\,\mathbf{E}_i\,\mathbf{p}^*\,,
\end{split}
\end{equation}
where we introduced the projection matrix $\mathbf{E}_i:=\hat{\mathbf{e}}_i\,\hat{\mathbf{e}}_i^{\text{T}}$.
Evaluating the double sum in Eq.~(\ref{eq:A2}) results in
\begin{equation}
\label{eq:A6}
p(\delta x,t;\delta x',t+\tau)=\boldsymbol{\eta}^{\text{T}}\,\boldsymbol{\Lambda}(\delta x')\,\mathbf{T}^{\tau}\,\boldsymbol{\Lambda}(\delta x)\,\mathbf{p}^*
\end{equation}
with $\boldsymbol{\eta}^{\text{T}}=(1,1,1,1)$ and $\boldsymbol{\Lambda}(\delta x)=\text{diag}(\lambda_1(\delta x),\lambda_2(\delta x),\lambda_3(\delta x),\lambda_4(\delta x))$.
Evaluating the double integral in Eq.~(\ref{eq:A1}), we obtain Eq.~(\ref{eq:jump_corr2}).
For the correlation function of the algebraic signs $S_t=\text{sign}(\delta x_t)$ of the spatial shifts, an analogous derivation can be done leading to Eq.~(\ref{eq:jump_corr1}).

\subsection{Derivation of Eq.~(\ref{eq:zig-zag_dist})}

In the following, we derive an analytical expression for the distribution of zig-zag streaks of length $\Delta t$ defined in the main text for an even value of $\Delta t$.
An analogous derivation can be done for an odd value of $\Delta t$.
Because of the symmetry of the distribution of spatial shifts, zig-zag streaks of length $\Delta t$ starting with $+1$ or $-1$ are equiprobable.
Therefore, the probability of a zig-zag streak of length $\Delta t$ is twice the probability of observing a sequence of algebraic signs $+1,-1,+1,\dots,+1,+1$ at discrete instants of time from $0$ to $\Delta t+1$,
\begin{equation}
\label{eq:B1}
p(\Delta t)=2\,p(+1,0;-1,1;+1,2;\dots;+1,\Delta t;+1,\Delta t+1)\,.
\end{equation}
By introducing a joint probability also containing the state $i_t$ of the HMM at discrete time $t$ and using the main property of Markov processes, we can write
\begin{equation}
\begin{split}
\label{eq:B2}
&p(\Delta t)=\\
&2\hspace{-0.6em}\sum\limits_{i_0,\dots,i_{\Delta t+1}=1,2,3,4}\hspace{-2.0em}p(+1,i_0;-1,i_1;+1,i_2;\dots;+1,i_{\Delta t};+1,i_{\Delta t+1})\\
&=2\hspace{-1.9em}\sum\limits_{i_0,\dots,i_{\Delta t+1}=1,2,3,4}\hspace{-2.0em}p(+1,i_{\Delta t+1}|+1,i_{\Delta t})\,p(+1,i_{\Delta t}|-1,i_{\Delta t-1})\\
&\hspace{5.0em}\cdots\,p(-1,i_1|+1,i_0)\,p(+1,i_0)\,.
\end{split}
\end{equation}
The conditional probability $p(+1,i_{\Delta t+1}|+1,i_{\Delta t})$ of being in state $i_{\Delta t+1}$ at time $\Delta t+1$ with emission $+1$
under the condition of being in state $i_{\Delta t}$ at time $\Delta t$ with emission $+1$ can be written as
\begin{equation}
\label{eq:B3}
p(+1,i_{\Delta t+1}|+1,i_{\Delta t})=\lambda_{i_{\Delta t+1}}^+\,\hat{\mathbf{e}}_{i_{\Delta t+1}}^{\text{T}}\,\mathbf{T}\,\hat{\mathbf{e}}_{i_{\Delta t}}
\end{equation}
with the same notation as in the previous appendix.
Similar expressions can be found for all conditional probabilities appearing in Eq.~(\ref{eq:B2}).
Under the assumption that the process has already converged to the stationary distribution $\mathbf{p}^*$, the probability $p(+1,i_0)$ of being in state $i_0$ at time $0$ with emission $+1$ can be written as
\begin{equation}
\label{eq:B4}
p(+1,i_0)=\lambda_{i_0}^+\,\hat{\mathbf{e}}_{i_0}^{\text{T}}\,\mathbf{p}^*\,.
\end{equation}
By inserting Eq.~(\ref{eq:B3}) and Eq.~(\ref{eq:B4}) into Eq.~(\ref{eq:B2}) and evaluating the ($\Delta t+2$)-fold sum in Eq.~(\ref{eq:B2}), we obtain Eq.~(\ref{eq:zig-zag_dist}).

\section{\label{sec:B}Determination of the parameters of the Hidden Markov Model for $\mu=-0.10$}

In order to compare the analytical results for the Hidden Markov Model (HMM) shown in Fig.~\ref{fig:hmm1} of the main text with the corresponding numerical results for the soliton motion,
we have to determine the parameters of the HMM, i.e., the values of the four probabilities $a$, $b$, $t_1$, and $t_2$ and the emission densities $\lambda_i(x)$, i.e.,
the probability densities on the four states from which the output on the corresponding state is drawn randomly.

We start with the analytical expression for the correlation function $C_S(\tau)$ of the algebraic signs of the spatial shifts in Eq.~(\ref{eq:jump_corr1}) of the main text.
We use the spectral decomposition of the transition matrix $\mathbf{T}$, i.e., $\mathbf{T}=\sum_{i=1}^4\sigma_i|\sigma_i\rangle\langle\sigma_i|$,
where $\sigma_i$ are the eigenvalues, and $|\sigma_i\rangle$ and $\langle\sigma_i|$ are the right and left eigenvectors, respectively, of the transition matrix $\mathbf{T}$
obeying $\langle\sigma_i|\sigma_j\rangle=\delta_{ij}$ and $\sum_{i=1}^4|\sigma_i\rangle\langle\sigma_i|=\mathbf{I}$, where $\mathbf{I}$ is the identity matrix.
It follows that $\mathbf{T}^{\tau}=\sum_{i=1}^4\sigma_i^{\tau}|\sigma_i\rangle\langle\sigma_i|$.
The eigenvalues of the transition matrix $\mathbf{T}$ in Eq.~(\ref{eq:transition_matrix}) are given by
\begin{equation}
\begin{split}
\label{eq:eigenvalues}
\sigma_1&=1,\quad\sigma_2=1-t_1-t_2,\\
\sigma_{3/4}&=\frac{1}{2}\biggl(-2+2a+2b+t_1+t_2\\
&\hspace{1.1em}\mp\left.\sqrt{(-2a+2b+t_1)^2+2(2a-2b+t_1)t_2+t_2^2}\right)\,.
\end{split}
\end{equation}
Interestingly, the dyadic product of right and left eigenvectors belonging to the first and second eigenvalue multiplied with the vectors and matrices appearing in Eq.~(\ref{eq:jump_corr1}) gives zero
as long as the symmetry $\lambda_1(-x)=\lambda_2(x)$ and $\lambda_3(-x)=\lambda_4(x)$ is fulfilled.
Therefore, the third and fourth eigenvalue of $\mathbf{T}$ to the power of $\tau$ are responsible for the bi-exponential decay of the correlation function $C_S(\tau)$.
Using the fitted bi-exponential decay of the correlation function $C_S(\tau)$ in Fig.~\ref{fig:statistics1} (a), it follows that $\sigma_3\approx-0.86$ and $\sigma_4\approx-0.20$.

We numerically found that the distribution $p(\Delta t)$ of zig-zag streaks of length $\Delta t$ in the soliton motion is very well described by a bi-exponential decay,
\begin{equation}
\label{eq:zig-zag_dist1}
p(\Delta t)\approx0.39\cdot e^{-0.56\cdot\Delta t}+0.013\cdot e^{-0.16\cdot\Delta t}\,.
\end{equation}
By using the geometric series, i.e., $\sum_{\Delta t=0}^{\infty}e^{-a\Delta t}=1/(1-e^{-a})$,
this normalized probability distribution with $\sum_{\Delta t=0}^{\infty}p(\Delta t)=1$ can be interpreted as the weighted sum of two normalized mono-exponential distributions,
\begin{equation}
\label{eq:zig-zag_dist2}
p(\Delta t)\approx0.91\cdot0.43\cdot e^{-0.56\cdot\Delta t}+0.09\cdot0.15\cdot e^{-0.16\cdot\Delta t}\,,
\end{equation}
where the weights are given by $0.91$ and $0.09$.
This weighted sum results from the weighted superposition of two Anti-Persistent Random Walks, where a single one shows a mono-exponential decay of the distribution $p(\Delta t)$.
In the HMM shown in Fig.~\ref{fig:hmm1} of the main text, state $1$ and $2$ belong to the first Anti-Persistent Random Walk (with persistence parameter $a$) and state $3$ and $4$ to the second one (with persistence parameter $b$),
where transitions between them are given by the probabilities $t_1$ and $t_2$.
The stationary distribution of the HMM results from the transition matrix $\mathbf{T}$ in Eq.~(\ref{eq:transition_matrix}) of the main text and is given by
\begin{equation}
\label{eq:stationary_distribution}
\mathbf{p}^*=\frac{1}{2(t_1+t_2)}\begin{pmatrix}t_1\\t_1\\t_2\\t_2\end{pmatrix}
\end{equation}
and should, therefore, reproduce the weights of the mono-exponential distributions, i.e., $t_1/(t_1+t_2)\approx0.91$ and $t_2/(t_1+t_2)\approx0.09$.
It follows that $t_1/t_2\approx10.11$.

Furthermore, very long zig-zag streaks in the soliton motion are purely caused by the Anti-Persistent Random Walk with the smaller persistence parameter.
In Sec.~\ref{sec:IV} of the main text, we numerically determined the distribution of zig-zag streaks for filtered sequences of spatial shifts $\delta x_t$.
For the filtered sequence which contains all $\delta x_t$ belonging to the sharp peaks in the distribution of spatial shifts shown in Fig.~\ref{fig:trajectories} (d),
we found a bi-exponential decay of the distribution of zig-zag streaks, where the second exponential decay is nearly identical to the second exponential decay of the distribution of zig-zag streaks for the unfiltered data.
This means that the Anti-Persistent Random Walk belonging to the states 3 and 4 of the HMM shown in Fig.~\ref{fig:hmm1} is responsible for the very long zig-zag streaks.
As a consequence and according to that HMM, the second exponential decay of the distribution of zig-zag streaks in Fig.~\ref{fig:statistics1} (b) should be proportional to $(1-b-t_1)^{\Delta t}$.
Therefore, we can conclude that $b+t_1\approx0.15$.

These considerations lead to the following system of equations
\begin{equation}
\label{eq:equations1_}
\sigma_3\approx-0.86,\,\sigma_4\approx-0.20,\,b+t_1\approx0.15,\,t_1/t_2\approx10.11,
\end{equation}
whose solution gives us the desired values of the parameters of the HMM,
\begin{equation}
\label{eq:parameters1_}
a\approx0.39,\quad b\approx0.001,\quad t_1\approx0.14,\quad t_2\approx0.014\,.
\end{equation}

We approximated the numerically determined distribution of spatial shifts shown in Fig.~\ref{fig:trajectories} (d) of the main text by a weighted sum of four Gaussian distributions, which is also shown in Fig.~\ref{fig:trajectories} (d),
\begin{equation}
\begin{split}
\label{eq:jump_dist1}
p(\delta x)\approx\sum\limits_{k=1}^4\omega_k\,\mathcal{N}(\delta x|u_k,\sigma_k^2)\,,
\end{split}
\end{equation}
where $\mathcal{N}(x|u,\sigma^2)$ is a Gaussian distribution with mean value $u$ and variance $\sigma^2$, i.e., $\mathcal{N}(x|u,\sigma^2)=1/(\sqrt{2\pi\sigma^2})\exp(-(x-u)^2/(2\sigma^2))$.
Due to symmetry, we have $u_1=-u_4$, $\sigma_1^2=\sigma_4^2$, and $\omega_1=\omega_4$ as well as $u_2=-u_3$, $\sigma_2^2=\sigma_3^2$, and $\omega_2=\omega_3$
with the numerically determined values $u_1=-2.16$, $u_2=-0.77$, $\sigma_1^2=0.026$, $\sigma_2^2=0.24$, $\omega_1=0.27$ and $\omega_2=0.23$ and this means that
the first and the fourth Gaussian distributions represent the sharper peaks of the distribution of spatial shifts and the second and third one represent the broader humps.

In order to reproduce this weighted sum of Gaussian distributions by the Hidden Markov Model, we choose Gaussian mixtures for the emission densities $\lambda_i(x)$ on state $i$.
In general, we can write
\begin{equation}
\label{eq:emissions}
\lambda_i(\delta x)=\sum\limits_{j=1}^2\phi_{ij}\,\mathcal{N}(\delta x|u_{ij},\sigma_{ij}^2)\,,
\end{equation}
where the index $j$ identifies the broader humps ($j=1$) and the sharp peaks ($j=2$) and one has $\phi_{i1}+\phi_{i2}=1$.
Because of the exact left-right symmetry of the system, we have for the parameters of the output distributions
$u_{1j}=-u_{2j}$, $\sigma_{1j}^2=\sigma_{2j}^2$, $\phi_{1j}=\phi_{2j}$, $u_{3j}=-u_{4j}$, $\sigma_{3j}^2=\sigma_{4j}^2$, and $\phi_{3j}=\phi_{4j}$,
and from the numerically observed ``top-down'' symmetry of the Gaussian shapes, we get $u_{1j}=u_{3j}$, $\sigma_{1j}^2=\sigma_{3j}^2$, $u_{2j}=u_{4j}$, and $\sigma_{2j}^2=\sigma_{4j}^2$.
Because of the distribution of zig-zag streaks for the filtered data shown in Fig.~\ref{fig:numerics_theory1} (a) of the main text,
which decays mono-exponentially for the broader humps but bi-exponentially for the sharper peaks, the latter must contribute to both Anti-Persistent Random Walks.
Therefore, we choose a Gaussian mixture for the emission densities $\lambda_{1/2}(x)$ of the HMM shown in Fig.~\ref{fig:hmm1} of the main text and a simple Gaussian distribution for the emission densities $\lambda_{3/4}(x)$, i.e.,
\begin{equation}
\begin{split}
\label{eq:emissions_}
\lambda_1(\delta x)&=\phi_{12}\,\mathcal{N}(\delta x|u_1,\sigma_1^2)+\phi_{11}\,\mathcal{N}(\delta x|u_2,\sigma_2^2)\,,\\
\lambda_2(\delta x)&=\phi_{11}\,\mathcal{N}(\delta x|-u_2,\sigma_2^2)+\phi_{12}\,\mathcal{N}(\delta x|-u_1,\sigma_1^2)\,,\\
\lambda_3(\delta x)&=\mathcal{N}(\delta x|u_1,\sigma_1^2)\,,\\
\lambda_4(\delta x)&=\mathcal{N}(\delta x|-u_1,\sigma_1^2)
\end{split}
\end{equation}
with $\phi_{11}+\phi_{12}=1$.
If the emission densities $\lambda_i(\delta x)$ are weighted by the stationary distribution of the HMM in Eq.~(\ref{eq:stationary_distribution}), they should reproduce Eq.~(\ref{eq:jump_dist1}), i.e.,
\begin{equation}
\label{eq:jump_dist2}
p(\delta x)\approx\sum\limits_{i=1}^4p_i^*\,\lambda_i(\delta x)\,.
\end{equation}
We can conclude that, e.g., $p_1^*\,\phi_{11}=\omega_2$ leading to the weights of the Gaussian mixture,
\begin{equation}
\label{eq:parameters2_}
\phi_{12}=0.49,\quad\phi_{11}=0.51\,.
\end{equation}

\begin{acknowledgments}
T.A. and G.R. gratefully acknowledge funding by the Deutsche Forschungsgemeinschaft (DFG, German Research Foundation) – RA 416/13-1.
J.C. thanks FONDECYT (Chile) for financial support through Grants 1210297 and 1200357.
\end{acknowledgments}

\bibliography{references}

\end{document}